\documentclass[a4paper]{JHEP3}

\pdfoutput=1

\usepackage[centertags]{amsmath}
\usepackage{amssymb,amstext,amsmath}
\usepackage{graphicx}
\usepackage{epsfig}
\usepackage[stable]{footmisc}
\usepackage{appendix}

\usepackage{young}
\usepackage{pbox}

\newcommand{\be}{\begin{equation}}
\newcommand{\ee}{\end{equation}}

\newcommand{\half}{\frac{1}{2}}
\newcommand{\thalf}{\frac{3}{2}}

\newcommand{\del}{\partial}
\newcommand{\lra}{\leftrightarrow}

\newcommand{\ra}{\rightarrow}
\newcommand{\orao}{\overrightarrow{0}}

\newcommand{\E}{\epsilon}

\newcommand{\B}{\beta}

\newcommand{\bpsi}{\bar{\psi}}
\newcommand{\tphi}{\tilde{\phi}}
\newcommand{\tpsi}{\tilde{\psi}}

\newcommand{\nytab}[2]{\ensuremath{   {\!}^{ #2 \left\{
  \begin{array}{l l}  \\    \\    \\ \end{array}\right. }
\overbrace{   \hspace{-4mm}\mbox{\begin{Young}
& & $\cdot \cdot \cdot$  & \cr
\cr
\vdots 
\cr
\end{Young}} }^{#1}      }}

\newcommand{\nytabbar}[2]{\ensuremath{        {\!}^{ #2 \left\{
  \begin{array}{l l}  \\    \\     \\  \end{array}\right. }
\overbrace{   \overline{ \hspace{-4mm}\mbox{\begin{Young}
& & $\cdot \cdot \cdot$  & \cr
\cr 
\vdots 
\cr
\end{Young}}} }^{#1}             }}

\newcommand{\nysymtab}[1]{\ensuremath{  \overbrace{ \begin{Young} & & $\cdot \cdot \cdot$  & \cr \end{Young} }^{#1} }}

\newcommand{\nysymtabbar}[1]{\ensuremath{
 \overbrace{ \overline{\begin{Young} & & $\cdot \cdot \cdot$  & \cr \end{Young}} }^{#1}
                      }}

\newcommand{\nyantisymtab}[1]{\ensuremath{
{\!}^{ #1 \left\{   \begin{array}{l l}  \\  \\   \\  \end{array}\right. } \hspace{-2mm}  
 \begin{Young} \cr \cr \vdots \cr \end{Young}   }}

\newcommand{\nyantisymtabbar}[1]{\ensuremath{
{\!}^{ #1  \left\{   \begin{array}{l l}  \\  \\   \\  \end{array}\right. } \hspace{-2mm}  
 \overline{\begin{Young} \cr \cr \vdots \cr \end{Young}}  }}

\preprint{WIS/10/14-DEC-DPPA}

\title{On monopole operators in supersymmetric Chern-Simons-matter theories}

\author{
Ofer Aharony, Prithvi Narayan and Tarun Sharma \\
\it{Department of Particle Physics and Astrophysics,\\
Weizmann Institute of Science, Rehovot 7610001, Israel}\\
E-mails : {\tt Ofer.Aharony@weizmann.ac.il, prithvi@weizmann.ac.il, Tarun.Sharma@weizmann.ac.il}
}

\abstract{We discuss monopole operators in $U(N_c)$ Chern-Simons-matter theories in three
space-time dimensions. We mention an apparent problem in the matching of such operators in
dualities between non-supersymmetric theories, and suggest a possible resolution. A similar
apparent problem exists in the mapping of chiral monopole operators in theories with ${\cal N}=2$
supersymmetry. We show that in many theories the lowest naive chiral monopole operator is
actually not chiral, and we find the lowest monopole operator that is actually chiral in these
theories. It turns out that there are several different forms of this operator, depending
on the number of colors, the number of flavours, and the Chern-Simons level. Since we use the
supersymmetric index to find the lowest chiral monopoles, our results for these monopoles
are guaranteed to be invariant under the dualities in supersymmetric theories. The theories
we discuss are believed to be dual in the 't~Hooft large $N_c$ limit to classical high-spin
gravity theories. We argue that these theories (supersymmetric or not) should not have
classical solutions charged under the $U(1)$ gauge field in the high-spin multiplet.}

\begin{document}

\section{Introduction}

In the last twenty years, many examples of dualities between different quantum field theories in three and
four space-time dimensions have been discovered. In particular, following \cite{Seiberg:1994pq}, 
many examples of pairs of theories that are the same at low energies have been found, both in three 
and in four space-time dimensions.

A particular class of interesting gauge theories in three space-time dimensions is $U(N_c)$ gauge
theories with matter fields in the fundamental representation and with a Chern-Simons (CS) coupling
for the gauge field\footnote{One reason that these theories are interesting is that in the 't~Hooft
large $N_c$ limit with a finite number of matter fields, they are believed \cite{Klebanov:2002ja,Sezgin:2003pt,Giombi:2009wh,Aharony:2011jz,Giombi:2011kc,Chang:2012kt} to be dual to classical
high-spin gravity theories on $AdS_4$ \cite{Vasiliev:1992av}.}. These theories can either be defined as the low-energy limit of gauge theories
which have both the Yang-Mills kinetic term and the Chern-Simons term (these theories can
flow to non-trivial conformal field theories at low energies if all relevant couplings are tuned to zero), 
or directly (without a
Yang-Mills term) as conformal field theories in which all beta functions vanish. In either case at
low energies the gauge field is not dynamical, but the matter fields are dynamical and their
couplings are affected by the Chern-Simons gauge fields. 

For theories of this type with ${\cal N}=2$ supersymmetry, dualities were discovered in 
\cite{Giveon:2008zn} for the case with $N_f$ chiral superfields in the fundamental representation, 
and $N_a=N_f$ chiral superfields in the anti-fundamental representation of $U(N_c)$ (this duality can be derived
by adding real mass terms to the duality without Chern-Simons coupling that was discovered
in \cite{Aharony:1997gp}). This was later generalized in \cite{Benini:2011mf} to the case with 
$N_a \neq N_f$.

Theories of this type without supersymmetry were studied in 
\cite{Aharony:2011jz,Giombi:2011kc}, and this led to a
conjecture that they also satisfy a duality between $U(N_c)_k$ theories ($k>0$) with $N_f$ scalar matter fields and
$U(k-N_c)_{-k+\frac{N_f}{2}}$ theories with $N_f$ fermion matter fields; this duality was presented explicitly in \cite{Aharony:2012nh} 
\footnote{Here we use the convention for $k$ that is natural from the low-energy point of view, as in \cite{Aharony:2012nh}.
In the non-supersymmetric theories this differs by a shift of $k$ by $N_c {\rm sign}(k)$ from the high-energy value of $k$ in Yang-Mills-Chern-Simons theories, so that we always have $|k| > N_c$.}. 
In the non-supersymmetric theories it is only known how to perform explicit computations at weak coupling or in the
large $N_c$ 't Hooft limit, so the evidence for the non-supersymmetric dualities at finite
$N_c$ is much weaker. It was shown in \cite{Jain:2013gza} that one can flow (at least for large enough
$N_c$) from the ${\cal N}=2$ dualities to the non-supersymmetric dualities, providing
evidence for the validity of the latter at finite $N_c$.

The statement of the duality is that in the low-energy conformal field theory (CFT), all operators
should match (including their scaling dimensions), and all their correlation functions as well.
In the supersymmetric case, it is possible to check that all chiral operators agree between
the two theories by computing their ``superconformal index'' 
\cite{Romelsberger:2005eg,Kinney:2005ej,Bhattacharya:2008zy} which is a sum over all chiral
operators. This index, proportional to the partition function on $S^2\times S^1$ with appropriate
background fields, can be computed \cite{Kim:2009wb,Imamura:2011su} using localization in terms 
of the high-energy degrees of freedom, and indeed in all cases that have been checked the index agrees
between pairs of dual theories \cite{Romelsberger:2007ec,Bashkirov:2011vy,Bashkirov:2011pt,Hwang:2011qt}. 
It is not known how to compare non-chiral 
operators (or any operators in the non-supersymmetric case at finite $N_c$), since it is not known
how to compute their dimension except at weak coupling.

In this paper we discuss ``monopole operators'' in these CFTs. A monopole operator 
is defined as a point-like defect such that there is some magnetic flux on the $S^2$
surrounding it (this flux can be chosen to be in the Cartan subalgebra of $U(N_c)$). 
It is related by the state-operator mapping to a state of the conformal field theory on
$S^2$ which has a non-zero gauge field flux on the $S^2$. In theories without Chern-Simons
couplings, such operators were studied extensively in the literature 
(see e.g. \cite{Borokhov:2002ib,Borokhov:2002cg,Pufu:2013eda,Pufu:2013vpa,Dyer:2013fja}). Their 
dimensions can be computed at weak coupling, but in non-supersymmetric theories
essentially nothing is known about them at higher
values of the coupling. In general gauge theories (say, with $SU(N_c)$ gauge group) it is not
even clear how to identify monopole operators at strong coupling. However, in $U(N_c)$ 
theories there is a `topological' $U(1)_J$  global symmetry 
whose current is the dual of the diagonal $U(1)$ field strength, and monopoles (and only
monopoles) are charged under this symmetry. This enables a simple identification of 
monopole operators even away from weak coupling.

In Chern-Simons theories there is an extra complication. The Chern-Simons term implies that
monopole operators carry an electric charge, so that to form a gauge-invariant
operator they must be dressed with extra charged fields. For example, the simplest monopoles
in the $U(N_c)_k$ theory break $U(N_c) \to U(1)\times U(N_c-1)$ and carry $\pm k$ units of charge
under the $U(1)$, and this charge must be balanced by extra fields carrying $\mp k$ units of
charge.

In a theory that contains scalar fields $\varphi$ in the fundamental and anti-fundamental representation,
one would expect the lightest (lowest dimension) monopole operator to arise from a product of
the monopole defect operator $X$ with $|k|$ scalar fields in the fundamental or anti-fundamental
representation (depending on the sign of $k$ and on the monopole charge), so that it takes the
form $X \varphi^{|k|}$. At weak coupling (large $|k|$) 
this naively gives an operator with a dimension of order $|k|$.  Note that in the theory on $S^2$
the lowest energy scalar states charged under the $U(1)$ have spin $\frac{1}{2}$ in the monopole
background \cite{Wu:1976ge}, so this product is actually not a scalar but an operator with spin $\frac{|k|}{2}$.

If there are no scalar
fields of the appropriate representation (which happens on one-side of the non-supersymmetric
duality) one needs to put in $|k|$ fermions $\psi$, but then because of anti-symmetry one needs to add
also $O(k^{\frac{3}{2}})$ derivatives in the large $|k|$ limit (see appendix \ref{fermion_counting}) to form a non-vanishing operator of the schematic form 
$(X \psi \del \psi \del^2 \psi \cdots)$. 
At weak coupling this
operator seems to have a dimension of order $|k|^{\frac{3}{2}}$. For the non-supersymmetric
duality one needs to map monopoles with scalars to monopoles with fermions, but this seems
problematic since their classical scaling dimensions are very different from each other, and even
scale differently with $N_c$ in the 't Hooft limit (in which one takes large $N_c$ with fixed
't Hooft coupling $\lambda \equiv N_c/k$). Recall that the dualities match the $U(1)_J$
symmetries on the two sides, so monopoles must map to monpoles under the duality. Presumably 
the monopoles acquire large anomalous dimensions at strong coupling that make this
matching work, but the needed anomalous dimensions do not satisfy the usual large $N_c$
scaling, and it would be interesting to understand where they come from.\footnote{The duality
actually maps the free scalar theory coupled to Chern-Simons to a Gross-Neveu model coupled
to Chern-Simons, but we do not expect going from the free theory to the critical one
to affect the large $N_c$ scaling of the monopole dimensions.}

In order to shed more light on this one can look at similar questions in ${\cal N}=2$ supersymmetric
theories, where at least for chiral monopole operators we have more control. In supersymmetric
theories such a monopole looks like a chiral field of the form (say) $V_+ \Phi^{|k|}$, where
$V_+$ is the standard chiral monopole operator with the minimal positive monopole charge 
(see \cite{Aharony:1997bx,Borokhov:2002ib,Borokhov:2002cg,Intriligator:2013lca}) and $\Phi$ is a chiral superfield in the
fundamental representation (if there are several such fields they could all appear). This seems 
to give a chiral operator of spin $\frac{|k|}{2}$. It is easy
to compute the classical dimension of this operator, and for the chiral operator one
expects this dimension to be protected.  Thus, naively we would expect the dimensions
of the $V_+ \Phi^{|k|}$ operators to match across supersymmetric dualities. However, it
is easy to see that because of the different scaling dimensions of the monopole operators
$V_+$ this is not the case, both in the dualities of \cite{Giveon:2008zn} and 
in the more general dualities of \cite{Benini:2011mf}.

How is this possible, given that the indices of the two theories, and thus all chiral operators,
match ? A deeper look at the index reveals that in many cases the operators $V_+ \Phi^{|k|}$
do not appear in the index (namely, there is no contribution to the index with the corresponding
quantum numbers), implying that they are actually not chiral. When a naively chiral operator
does not appear in the index, this means that it can join with another operator to form
a non-chiral multiplet, and generically we expect this to happen whenever it can. As we discuss in detail
below, from the point of view of the index which is computed in the UV theory, there
are other operators with the same quantum numbers as $V_+ \Phi^{|k|}$ that involve gluinos, which can join
together with these operators to form long (non-chiral) multiplets of the superconformal
algebra. From the point of view of the low-energy theory that contains no gluinos (and no other
naively chiral operators with the same quantum numbers) this non-chirality
is more surprising, but this theory is generally strongly coupled.

In this paper we study in detail the spectrum of chiral monopole operators in $U(N_c)_k$
theories, focusing for simplicity on the two cases $N_a=N_f$ and $N_a=0$. The latter
case is particularly interesting because it is used (for $N_f=1$) to flow to the 
non-supersymmetric duality \cite{Jain:2013gza}.
We will show that in some cases the naive chiral operators are chiral, but in other cases they
are not, and for every value of $N_c, N_f$ and $k$ we identify the lightest monopole
operator that appears in the index. Our results are based partly on a numerical
evaluation of the index for small values of $N_c, N_f$ and $k$, which we use to
conjecture the general result, and partly on analytic arguments that are valid for some ranges of values
of $N_c, N_f$ and $k$. We verify that these lightest operators match across
the duality, as implied by the equality of the indices of dual theories. Note that in
general the lightest monopole operators carry a non-zero spin.

In some cases we find
that the lightest chiral monopole operator has a dimension of order $N_c^2$ in the 't Hooft 
large $N_c$ limit. This implies that all the naive chiral monopole operators with dimensions 
of order $N_c$ are actually not chiral. 
We then go back to the apparent mismatch in non-supersymmetric theories, and argue that
already at weak coupling in the scalar theory, the monopole operators could get large
anomalous dimensions that may change their $N_c$-scaling in the 't Hooft large $N_c$ limit.

We begin in section \ref{background} with a review of background material about monopole
operators, superconformal indices, and how to read off the charges and field content of
chiral monopole operators from the index. In section \ref{nonchiral} we present a conjecture
(based on numerical evaluations of the index) for the
dimensions and flavour representations of the lowest monopole operators for the case of
$N_a=N_f$, and in section \ref{chiral} we do the same for $N_a=0$. In section
\ref{LNTL} we discuss the 't~Hooft large $N_c$ limit of our results. We use the duality
between CS-matter theories in this limit and high-spin gravity theories to argue that
the latter theories should not have classical charged solutions. In section \ref{analyticone}
we prove our conjecture for the form of the lowest chiral monopole operator in
a simple case; other cases are analyzed in appendix \ref{apdx:AA}. In
section \ref{nonchiraldual} we briefly discuss the duals of non-chiral monopole operators
under the supersymmetric dualities.
In section \ref{nspc}
we discuss the perturbative corrections to dimensions of monopole operators
in non-supersymmetric Chern-Simons-matter theories, and argue that they can be as
large as $O(k^{\frac{3}{2}})$. We summarize our results in section \ref{summary}. Several appendices contain technical details.

\section{The Superconformal Index and BPS monopole operators}
\label{background}

The Superconformal Index $I$ of a 3d ${\cal N}=2$ supersymmetric theory 
\cite{Bhattacharya:2008zy,Kim:2009wb,Imamura:2011su,Krattenthaler:2011da,Kapustin:2011jm}
is defined as a weighted sum over the Hilbert space of the theory on $S^2$ as follows:
\be 
I={\rm Tr} \left[(-1)^F e^{-\B\{Q,S\}}x^{\epsilon+j_3}\prod_n t_n^{f_n}\right],
\ee
where 
\begin{itemize}
\item $F$ is a fermion number operator and $(-1)^F$ gives $(+1)$ for bosonic and $(-1)$ for fermionic 
states.
\item $Q$ and $S$ are particular supercharges in the superconformal algebra which 
satisfy 
\be \label{zero}
\{Q,S\}=\epsilon-j_3-R \geq 0,
\ee
where $\epsilon$ is the energy in units of the radius of the $S^2$, $j_3$ is the charge under the Cartan subalgebra of the
$Spin(3)$ rotation group of the $S^2$, and $R$ is the R-charge of the ${\cal N}=2$ 
superconformal algebra. Under radial quantization $Q$ are $S$ are Hermitian conjugates of 
each other and \eqref{zero} is positive semi definite.
\item Only states with $\{Q,S\}=0$ contribute to $I$, so it is actually independent of $\beta$.
\item $I$ is a non-trivial function of $x$ and of all the other fugacities for global symmetries $t_n$,
and $f_n$ are the charges under these symmetries.
\item $I$ is invariant under continuous deformations of the theory which preserve the 
superconformal symmetry of the theory. 
\end{itemize}

Up to an overall factor related to the vacuum energy of the theory on $S^2$,
$I$ is equal to the partition function of the theory on $S^2\times S^1$ with appropriate
background fields. Thus, it
can be evaluated by a path integral of the theory on $S^2\times S^1$,
with periodic boundary conditions for both 
fermions and bosons, and with the relevant chemical potentials turned on. This is a 
supersymmetric quantity and can be evaluated via supersymmetric localization 
on $S^2\times S^1$.  Since the index does not change under renormalization
group flow (except possibly for changes in the R-symmetry that sits in the
superconformal algebra), it can be defined even for theories which are not
conformal, as long as they have an exact $U(1)_R$ symmetry, and in
asymptotically free theories it can be easily computed in the UV. In particular,
in our case, it can computed in a Yang-Mills-Chern-Simons
theory that is weakly coupled at high energies.

In this paper we are interested in $U(N_c)$ CS theories at level $k$ 
coupled to $N_f$ chiral multiplets $\Phi_a$ in the fundamental and $N_a$ chiral multiplets ${\tilde \Phi}_{\tilde b}$ in the
anti-fundamental representation of $U(N_c)$ \footnote{For convenience we will use 
the shorthand notation $U(N_c)_k (N_f,N_a)$ for these theories.}. The flavour symmetry 
group of these theories is given by $(U(N_f)\times U(N_a))/U(1)$, with a combination of the 
two $U(1)$'s being a part of the gauge symmetry. We will take $\Phi_a$ (${\tilde \Phi}_{\tilde b}$) to be in the fundamental representation of $U(N_f)$ ($U(N_a)$). It will be convenient to write the fugacities of the global symmetry 
$SU(N_f) \times SU(N_a) \times U(1)_A$ as $t_a, \tilde t_{\tilde b}, y$, respectively, satisfying $\prod_{a=1}^{N_f} t_a = \prod_{\tilde b=1}^{N_a} \tilde t_{\tilde b}  = 1 $. In the special case $N_f=0$ or $N_a = 0$  there is no
$U(1)_A$ symmetry, so one must set $y=1$. There is also a topological $U(1)_J$ symmetry, whose
current includes $\epsilon^{\mu \nu \rho} {\rm tr}(F_{\nu \rho})$, and
whose fugacity we denote by $w$.  With these definitions, the superconformal index takes 
the following explicit form:
\be\begin{split}\label{SCIdef1}
I &=\sum_{\{m_i\} \in \mathbb{Z}} (-1)^{\sum (-k m_i-\half(N_f-N_a)|m_i|)} \frac{w^{\sum m_i}}{(sym)}  
       \oint \left( \prod_{i=1}^{N_c}  \frac{dz_i}{2\pi iz_i} z_i^{-km_i} \right)
    Z_g \left(\prod_{a=1}^{N_f}Z_{\Phi_a}\right) 
     \left(\prod_{{\tilde b}=1}^{N_a}Z_{\tilde{\Phi}_{\tilde b}}\right),\\
Z_g &= \prod_{(i\neq j)=1}^{N_c} x^{-|m_i-m_j|/2}\left(1-\frac{z_i}{z_j}x^{|m_i-m_j|}\right), \\
Z_{\Phi_a} &= \prod_{i=1}^{N_c} \left((x^{1-r}z_i^{-1}t_a^{-1}y^{-1})^{|m_i|/2}
   \prod_{j=0}^{\infty} \frac{(1-z_i^{-1}t_a^{-1} y^{-1}x^{|m_i|+2-r+2j})}{(1-z_i~t_a~y~x^{|m_i|+r+2j})}
   \right), \\
Z_{\tilde{\Phi}_{\tilde b}} &= \prod_{i=1}^{N_c} \left((x^{1-{\tilde r}}z_i \tilde{t}_{\tilde b}^{-1} y^{-1})^{|m_i|/2}
   \prod_{j=0}^{\infty} \frac{(1-z_i\tilde{t}_{\tilde b}^{-1}y^{-1}x^{|m_i|+2-{\tilde r}+2j})}
   {(1-z_i^{-1}\tilde{t}_{\tilde b}~y~x^{|m_i|+{\tilde r}+2j})}
   \right),
\end{split}\ee
where $(sym)$ is the dimension of the subgroup of the $S_{N_c}$ Weyl group that is unbroken by the
monopole background with fluxes $\{m_i\}$ $(i=1,\cdots,N_c)$ on $S^2$ in the Cartan of $U(N_c)$, $r$ is the 
R-charge of the $\Phi_a$, and $\tilde r$ of the ${\tilde \Phi}_{\tilde b}$ (these charges may be
modified by mixing them with other global symmetry charges, using the appropriate fugacities \footnote{The
precise R-symmetry of the superconformal theory in the IR can in principle be found by F-maximization \cite{Jafferis:2010un,Jafferis:2011zi,Closset:2012vg,Safdi:2012re}, but this
will not play any role in our analysis. Note that the IR R-symmetry can contain also accidental symmetries that
are not captured by the index \cite{Safdi:2012re}, but the index must still match between IR-dual theories.}). We include the phase factor $(-1)^{\sum (-km_i-\half(N_f-N_a)|m_i|)}$ which was 
pointed out in \cite{Dimofte:2011py} and which plays a crucial role in the factorization properties 
of the index studied in \cite{Hwang:2012jh,Benini:2013yva}.

As the Chern-Simons-matter theory we are studying is superconformal \cite{Gaiotto:2007qi}, there is a one-to-one map between local operators
on ${\mathbb R}^3$ and states on $S^2 \times {\mathbb R}$. 
In the sector with fluxes $\{m_i\}$ on $S^2$, the $U(N_c)$ gauge symmetry of the theory is broken 
to a subgroup which keeps the flux invariant. The flux state on $S^2$, which carries gauge charge $\{-km_i\}$ due
to the CS coupling, is dual to a local operator on ${\mathbb R}^3$ 
which is charged under the unbroken gauge symmetry. This is referred to as the `{\it bare}' monopole 
operator. This operator can be dressed with charged fields to make gauge-invariant monopole operators.

It is useful to keep track of the basic fields (`letters') which have the correct charges to contribute to the index. When we compute the contribution to the index from a
sector with fluxes $\{m_i\}$, we need to take into account how this shifts the quantum numbers of the various fields; we determine this from the
states of each field in the monopole background on $S^2$, by using the state/operator correspondence. Note that here we need to use the supersymmetric monopole background, which involves also 
an expectation value for the scalar field in the vector multiplet \cite{Borokhov:2002cg}. We denote the scalar and fermion components
of $\Phi_a$ (${\tilde \Phi}_{\tilde b}$) by $\phi_a$ and $\psi_a$ (${\tilde \phi}_{\tilde b}$ and ${\tilde \psi}_{\tilde b}$), respectively. The quantum numbers $(\epsilon,j_3,R;A)$ of the basic letters 
with $R$-charge $R$ and $U(1)_A$ charge $A$, in the flux background $\{m_i\}$, are given by:
\be\begin{split}
\label{letters}
\phi_i^a  &: \left( r+\half|m_i|,\half|m_i|,r;1 \right), \\
\tilde{\phi}^i_{\tilde b} &: \left( {\tilde r}+\half|m_i|,\half|m_i|,{\tilde r};1 \right), \\
\bar{\tilde{\psi}}_{+i}^{\tilde b} &: 
               \left( \thalf-{\tilde r}+\half|m_i|,\half(1+|m_i|),1-{\tilde r};-1 \right), \\
\bar{\psi}_{+a}^i &: 
               \left( \thalf-r+\half|m_i|,\half(1+|m_i|),1-r;-1 \right), \\
(\lambda_-)^i_j {\rm\ with\ } m_i \neq m_j &: \left(\frac{1}{2}(|m_i-m_j|+1),\frac{1}{2}(|m_i-m_j|-1),1;0 \right), \\
\del_{++} &: \left(1,1,0;0\right),
\end{split}\ee
where $i,j=1,\cdots,N_c$, and $\lambda^i_j$ are the gauginos of the high-energy Yang-Mills-Chern-Simons theory, which contribute only for $m_i \neq m_j$. The $\pm$'s denote which component we are considering, according to its charge under the Cartan subalgebra of the $Spin(3)$ rotation group, before taking into account the shift by the monopole background. Here we only wrote
down the `letters' which obey an equality in \eqref{zero}, since others do not contribute to the index;
whenever we write down a field contributing to an operator from here on, we will mean the
specific component of the field which is listed in \eqref{letters}. Operators containing
(say) $(\del^j \phi_i^a)^n$ are identified by having a contribution to the index from the
appropriate term in the Taylor expansion of the denominator of $Z_{\Phi_a}$,
and $\del^j {\bar \psi}_a^i$'s by contributions from the numerator \footnote{When we write down operators with
derivatives, they should always be interepreted as gauge-covariant derivatives.}.
Note that the gauginos
are not expected to be part of the low-energy CS-matter theory, so the interpretation of their
contributions from the point of view of this theory is not clear \footnote{This is possible since this theory
has non-zero couplings. The gauginos become auxiliary fields at low energies, that can be written as combinations
of the basic fields of the schematic form $\phi \bar\psi$.
These combinations differ from the naively chiral combinations of $\phi$ and $\bar\psi$ from \eqref{letters} that contribute
independently to the index, but
apparently (for $m_i \neq m_j$) they are chiral operators in the low-energy CS-matter theory.}, but we will
still write our operators using these gauginos (since this is how we can identify their form from the index, by
contributions from $Z_g$).

The quantum numbers of the bare chiral monopole operator $V_{\{m_i\}}$ can be computed as in \cite{Aharony:1997bx}, 
and they can also be read off from the Index.  
This operator is rotationally symmetric $(j_3=0)$, invariant under the $SU(N_f)\times SU(N_a)$ 
flavour symmetry, and it carries R-charge and axial charge\footnote{The axial $U(1)$ is a symmetry only 
when both $N_f$ and $N_a$ are non zero.} 
\be\begin{split}
\epsilon(V_{\{m_i\}}) =R(V_{\{m_i\}}) &= -\sum_{(i\neq j)=1}^{N_c} \frac{|m_i-m_j|}{2} 
                 + (N_f (1-r) + N_a(1-{\tilde r})) \sum_{i=1}^{N_c}\frac{|m_i|}{2},\\
A(V_{\{m_i\}}) &= (-1)(N_f+N_a) \sum_{i=1}^{N_c} \frac{|m_i|}{2}.  \\
\end{split}\ee

For simplicity let us first consider $U(N_c)_k(N_f,N_f)$ theories, in which ${\tilde r} = r$. 
Naively, for $k>0$ the lightest BPS monopole operator (namely, the one with the lowest
value of $\epsilon+j_3$) with unit topological charge is then
\footnote{Monopole operators from other GNO sectors turn out to be heavier.}
\be\label{nlV+}
{\cal M}_{(1,\orao)} = V_{(1,\orao)} (\phi_1)^k, {\rm \ contributing\ a\ factor\ } x^{N_f-N_c+1+k+r(k-N_f)}
\ee 
for all choices of flavours of the $\phi_1$ operators.
It turns out that this is not always the case. As we will discuss in later sections, 
in many cases this operator cancels with other operators in the index, and is 
thus presumably  not chiral. One can also see that this has to be the case from the Giveon-Kutasov (GK) 
duality in these theories \cite{Giveon:2008zn}. The GK duality relates the $U(N_c)_k(N_f,N_f)$ theory with 
$R_\phi=r$ and $A_\phi=1$ to a $U(|k|+N_f-N_c)_{-k}(N_f,N_f)$ theory with $R_\phi=1-r$ 
and $A_\phi=-1$ and with $N_f^2$ extra gauge-singlet chiral superfields $M$.
For consideration of the lightest monopole operator, $M$ is irrelevant. This implies 
that if \eqref{nlV+} is always the lightest monopole operator then its index 
contribution must match with the GK dual. This is easily seen not to be the 
case. 

The above observation implies that in some theories the naive leading chiral monopole 
operator must cancel in the index (since if not it would be below the lightest monopole
operator in the dual theory), and is thus {\it not} chiral.
This leads to the following 
questions:
\begin{itemize}
\item When does the naive leading monopole operator \eqref{nlV+} survive ? 
\item What is the leading non-canceling monopole operator in the cases when \eqref{nlV+} is not chiral ?
\end{itemize}
In the following sections we will answer these questions in detail, and further verify that 
the leading operators match across the GK duality.

Before proceeding to our computations, we would like to make a remark on different ways of 
computing the integral in \eqref{SCIdef1}. The index is represented as a contour integral over 
the unit circle in the complex plane for the holonomy variables $z_i$. The integrand has an infinite number 
of simple poles coming from the contributions $Z_{\Phi_a}$, $Z_{{\tilde \Phi}_{\tilde b}}$ of fundamental charged letters and their derivatives. Apart 
from these poles there are poles at $z_i=0$ or $\infty$ as well, depending on the value 
of $N_f$, $N_a$ and $k$. Naively the integral can be evaluated by summing over the residues at these poles,
and this has been done for $U(N_c)_k (N_f,N_a)$ theories in \cite{Hwang:2012jh}. However, in many cases
this procedure does not work (see the revised version of \cite{Benini:2013yva}), since it is not clear precisely how
to take into account the poles at $z_i=0$ and $z_i=\infty$. Thus, we will not use this method.
Instead we will evaluate the Index by performing 
a Laurent expansion of the integrand for small chemical potential $x$.

\section{Conjecture for the leading chiral monopoles in the $\sum_i m_i=1$ sector of  $U(N_c)_k (N_f,N_f)$ theories}
\label{nonchiral}

In this section we compute the lowest monopole operator (namely, the one with least $\epsilon + j_3$) 
which survives in the index of ``non-chiral'' theories (with $N_f = N_{a}$), in the sector with charge $+1$
under $U(1)_J$ (namely, $\sum_i m_i = 1$). We do this by expanding the 
Index \eqref{SCIdef1} order by order in $x$ using Mathematica and identifying the 
lowest power of $x$ which survives. The results of  Mathematica  suggest that the 
lowest nontrivial operator occurs (as expected) in the sector with GNO charges $(1,0,0,\cdots)$. Note that
the global symmetries do not distinguish sectors with different GNO charges $\{m_i\}$ and the same $U(1)_J$
charge $\sum_i m_i$, and these can mix (even though they appear separately in \eqref{SCIdef1}). For simplicity we give 
the results for $k>0$, from which the result for $k<0$ can easily be obtained, as will be explained in 
section \ref{consistencynonchiral}.

These operators all come with a factor of $w^{+1}$. The other global symmetries are $SU(N_f)_l \times SU(N_f)_r$ (whose fugacities 
are $(t_1,\ldots t_{N_f})$ and $(\tilde{t}_1,\ldots \tilde{t}_{N_f})$, respectively)
\footnote{With the condition $\prod_{a=1}^{N_f}t_a=1=\prod_{a=1}^{N_f}\tilde{t}_a$.}, along 
with the axial $U(1)_A$ symmetry (whose fugacity is $y$). 

Using the results above, the $x$ power, gauge charge and flavour charges of some basic relevant objects are \footnote{From here on we use the
shorthand notation $V_{\pm} \equiv V_{(\pm 1,\orao)}$ for the simplest monopole operator. This monopole breaks $U(N_c) \to U(1)\times U(N_c-1)$, and from here on sums over $i$ run over the $U(N_c-1)$ index, from $2$ to $N_c$.}
\be\begin{split}
V_+ &\ra z_1^{-k} x^{N_f-N_c+1-r N_f} y^{-N_f},
  \\
\{\phi_1^a, ~\lambda_1^i \phi_i^a \} 
&\ra z_1  x^{1+r} y ~t_a.\\
\end{split}\ee
One can easily see that the lightest gauge-neutral monopole operator 
is obtained by dressing $V_+$ by $k$ $\phi_1$'s.  However, in a somewhat surprising result, we find 
that this operator does not usually survive in the index, because the operators $(\lambda_1^i \phi_i^a)$
have the same quantum numbers as $\phi_1^a$ and come with an opposite sign. In fact, we find four regimes of 
$N_c,k,N_f$ (called Cases $1,2,3,4$) where we find different monopole operators giving the leading contribution
to the index. 

Table \ref{NonChiral} gives the results found using a numerical evaluation of the Index in Mathematica (extrapolated
from small
values of $N_c, N_f$ and $k$), where 
$N_c^d \equiv |k| +N_f - N_c$. In some cases we can confirm these results by analytic methods, as described
below. The charges of the lowest monopole operators which 
survive\footnote{One can verify that the lowest surviving operator is independent 
of the choice of $0<r<1$.} in various regimes are listed in the third column, along with a 
typical operator (there are generally other operators with the same charges, this is 
just a representative). The flavour representation of the leading surviving operator is 
given in the last column. Note that in all cases the results are continuous at the boundaries
of the different regimes, $k=N_c$ and $N_f=N_c$.

\begin{table}
\centering
\begin{tabular}{|c|c|c|c|}
\hline
& Region & Operator: $w^{\sum_i m_i} x^{\epsilon+j_3} y^A$ & $SU(N_f)_l\times SU(N_f)_r$ flavour rep. \\
\hline
1 & $N_f > N_c , k \geq N_c $ 
& \vspace{0cm}$V_+ \phi_1^{k-N_c+1} (\lambda_1^i \phi_i)^{N_c-1}$~: & ( $\nytab{k -N_c+1}{N_c}$ , 1) \\
&  & $w x^{ N^d_c+1 + r(k-N_f) } y^{k-N_f}$ &  \\
\hline
2 & $k \ge N_c \ge N_f$ 
& $V_+\phi_1^{k-N_c+1} (\lambda_1^i \phi_i)^{N_f-1} ({\lambda_1^i \tilde {\bar \psi}_i})^{N_c-N_f}$~: & $(\ \nysymtab{k-N_c}\ , \nysymtabbar{N_c-N_f}\ )$ \\
& & $w x^{k-N_f+N_c+1+r(N_c^d - N_c)} y^{N_c^d - N_c} \vspace{2mm}$ & \\
\hline
3 & $N_c \ge N_f , N_c >  k$ & $V_+\phi_1 (\lambda_1^i \phi_i)^{N_c^d-1} (\lambda_1^i {\tilde {\bar \psi}}_i)^{N_c-N_f} (\phi_i \tilde {\phi}^i)^{N_c-k} $:  & $(1, \nytabbar{N_c-N_f+1}{k+N_f-N_c})$ \\
& &  $w x^{k-N_f+N_c+1+r(N_f-k)} y^{N_f-k}$ & \\
\hline
4 & $ N_f \ge N_c \ge k $ 
& $V_+(\lambda_1^i \phi_i)^k (\phi_i \tilde {\phi}^i)^{N_c-k}$: & $ ( \nyantisymtab{N_c}\ \  , \nyantisymtab{N_c-k} \ )$  \\
& & $w x^{N^d_c+1+r (N_c- N_c^d)} y^{N_c- N_c^d}$ & \\
\hline
\end{tabular}
\caption{The leading ${\cal M}_+$ monopole operator in $U(N_c)_k(N_f,N_f)$ theories in various parameter ranges.}
\label{NonChiral}
\end{table}

Some more details are given in Appendix \ref{apdx:AA}, where we compute the lowest monopole operator in the Index analytically for several cases. The spins and R-charges of these monopoles are listed in Appendix \ref{apdx:DCM}.


\subsection{Consistency with duality}\label{consistencynonchiral}

Given the result for non-chiral  $U(N_c)_k$ theories with $k>0$, it 
is easy to read off the results for $k<0$. Let us denote the Index contribution in the 
GNO sector $\{m_i\}$ of the $U(N_c)_k(N_f,N_f)$ theory with $R_\Phi=R_{\tilde{\Phi}}=r$ as  
$I_{N_c,k,N_f,N_f}^{\{m_i\}}(r;x,y,w,t_a,\tilde t_a)$. From \eqref{SCIdef1} it is easy to 
see that 
\begin{equation}
I_{N_c,-k,N_f,N_f}^{\{m_i\}}(r;x,y,w,t_a,\tilde t_a) 
      =  I_{N_c,k,N_f,N_f}^{\{m_i\}}(r;x,y,w,\tilde t_a,t_a). 
\end{equation}
Therefore for $U(N_c)_{-k}$ theories, the same operator survives as in Table 
\ref{NonChiral}, except that the flavour Young tableaux are interchanged.  Note also that the Index of $U(N_c)_{-k}$ in the
$\{m_i\}$ sector is the same as the Index of $U(N_c)_k$ in the $\{-m_i\}$ sector, except for the power of $w$ (so the full indices are
the same up to $w \leftrightarrow w^{-1}$).

As a consistency check, we can now confirm that the results we obtained 
are consistent with the GK duality. Under the duality, 
Case 2 and Case 4 map into themselves, and Case 1 and 
Case 3 map into each other. This is consistent with the observation that specific monopole operators like
$V_+ \phi_1^k$ do not map to themselves under the duality.
%
%
The results in Table \ref{NonChiral} for the lowest surviving monopole operators 
are consistent with the GK duality, along with $r \ra 1-r, ~y \ra y^{-1}$, accompanied 
by complex-conjugating the flavour representation, as expected. More precisely, the lowest 
surviving operator is consistent with the duality relation
\be\begin{split}
\label{nonchiralduality}
I_{N_c,k,N_f,N_f}(r;x,y,w,t_a,\tilde t_a) = & I_{N_c^d,-k,N_f,N_f}(1-r;x,y^{-1},w, t^{-1}_a,\tilde t^{-1}_a) \\
= & I_{N_c^d,k,N_f,N_f}(1-r;x,y^{-1},w,\tilde t^{-1}_a,t^{-1}_a) \\
\end{split}\ee
(this is not a precise equality in general because of the extra singlet meson
operators that need to be added on the right-hand side).

$k=0$ is a special case. In this case the bare monopole operators 
$V_\pm$ are gauge-invariant by themselves and need not be dressed by charged matter 
fields. Since all other operators in the $\sum m_i=\pm 1$ sectors have larger values of 
$\epsilon+j_3$, $V_\pm$ itself has the lowest non-canceling contribution in 
the index. The matching of the lightest monopole across the duality 
also works differently in this case. The dual theories contain extra singlet chiral superfields  
$V_\pm$ charged under $U(1)_J$, along with superpotential terms  for their monopole operators $\tilde{V}_{\pm}$
\begin{equation}
W=\tilde{V}_+ V_- + \tilde{V}_- V_+.
\end{equation}
These superpotential terms make the bare monopole operators ${\tilde V}_{\pm}$ of the dual theory 
$Q$-exact and remove them from the chiral spectrum, while the gauge singlets 
$V_\pm$ map to $V_\pm$ of the original theory \cite{Aharony:1997gp}.

\subsection{Other GNO sectors}

In previous sections we presented the lowest monopole operator in the sector with GNO charge $(1,0,\cdots)$. One might be worried whether sectors with different GNO charges but the same topological charge $\sum_i m_i =1$ (say $(2,-1,0,\cdots)$) could give rise to a lower monopole operator, or cancel the contributions of the monopoles we presented. Using our numeric code (for low values of $N_c, N_f, k$ and the GNO charges) we explicitly checked that this is not the case.  The fact that the duality is consistent with the results of the previous subsection also suggests that this is not the case. 

In the simplest case  (case 1), when the lowest surviving monopole operator has the same power of $x$ as  $V_+ \phi_1^k$, we can explicitly show that this is indeed the lowest monopole operator with $\sum_i m_i = 1$. Consider a monopole operator with GNO charge $\{ m_i \}$, where $m_i$ are ordered by $|m_1| \ge |m_2| \ge \cdots$.  The naive gauge-invariant monopole operator\footnote{Note that this might not always survive in the index. Also assume $k>0$ for simplicity.} in this case is 
\begin{equation}
V_{\{ m_i\} } \left( \phi_1^{m_1}\cdots \phi_{N_c}^{m_{N_c}} \right)^k  \quad : \quad \epsilon+j_3 = - \sum_{i \ne j} {|m_i - m_j| \over 2} + N_f(1-r) \sum_i |m_i| + k \sum_i |m_i| (r + |m_i|),
\end{equation}  
where for any $m_i < 0$, we should use $ (\tilde \phi^i)^{k m_i}$ instead of $\phi_i^{k m_i}$. Using the triangle inequality $|m_i + m_j| \le |m_i| + |m_j|$, we find that for this operator
\begin{equation} \label{gen_mon}
\epsilon + j_3 \ge \sum_i |m_i| (N_c^d + 1 + (k- N_f) r  ) + k \sum_i |m_i| (|m_i| -1).
\end{equation}
It is now obvious that the only way we can minimize this charge keeping $\sum m_i = 1$ for arbitrary $r$ charge is to choose $m_1 =1$ and $m_i =0 $ for $i > 1$ (in this case there is equality in \eqref{gen_mon}).


\section{Conjecture for the leading chiral monopoles in the $\sum_i m_i= \pm 1$ sectors of  $U(N_c)_k (N_f,0)$ theories}
\label{chiral}

Let us now turn to a chiral case, $U(N_c)_k$ with $(N_f,0)$ matter fields. Note that one has to set $y=1$ in the general formula \eqref{SCIdef1}, since the axial symmetry is part of the gauge symmetry in this case. One can again perform a series of computations (similar to the non-chiral case) to find the lowest monopole operators in the sectors $\sum_i m_i  = \pm  1$.  We find that the lowest operator occurs at GNO charge $(\pm 1,0,\ldots)$.  Note that in this case there is no relation between the operators with $\{m_i\}$ and with $\{-m_i\}$ for the same value of $k$.
In particular, unlike the non-chiral case, the monopole operators with $\sum_i m_i=1$ are very different from $\sum_i m_i=-1$. 

Another important feature of these theories is that the duality works differently depending on the sign of $(k-\half N_f)$ \cite{Benini:2011mf}:
\be\begin{split}\label{csdtab}
& k-\half N_f \geq 0 : \quad U(N_c)_k (N_f,0) \overset{\text{dual}}{\leftrightarrows}
                          U(|k|+\half N_f-N_c)_{-k}(0,N_f), \\
& k-\half N_f \leq 0 : \quad U(N_c)_k (N_f,0) \overset{\text{dual}}{\leftrightarrows}
                          U(N_f-N_c)_{-k}(0,N_f), \\
\end{split}\ee
where for $k \neq \frac{N_f}{2}$ no extra singlet operators are needed for the duality.
We will call the two cases in \eqref{csdtab} Case 1 and Case 2, respectively. It will turn out that each of these cases has further subcases, where  the lowest monopole operator surviving in the index has a different form.


\subsection{Monopole GNO charge $(1,0,\ldots)$ sector}

The charges of the bare monopole $V_+$ correspond to a contribution to the index of the form
\begin{equation}
V_+ \rightarrow z_1^{-k-{N_f \over 2}} x^{-(N_c-1)+ (1-r){N_f \over 2}}.
\end{equation}
In Table \ref{Chiralplus} we give the results of Mathematica for the lowest monopole 
operator appearing in the index (again, these results are based on extrapolating numerical evaluations
of the index for small values of $N_c, N_f$ and $k$, and in some cases they can be
verified by analytic arguments).

\begin{table}
\centering
\begin{tabular}{|c|c|c|c|}
\hline
 & Region & Operator: $w^{\sum m_i} x^{\epsilon+j_3}$ & $SU(N_f)$ flavour rep. \\
\hline
1a & $(k>\half N_f, k+\half N_f>N_c);$
 & $V_+ \phi_1^{k+ {N_f \over 2}-N_c+1} (\lambda_1^i \phi_i)^{N_c-1} $: & $\nytab{k + {N_f \over 2}-N_c+1}{N_c}$ \\
 & $N_f\ge N_c$ & $w x^{k+N_f-N_c+1+kr}$ &  \\
\hline
1b & $(k>\half N_f, k+\half N_f>N_c);$ 
& $V_+\phi_1^{k-\half N_f+1}(\lambda_1^i\phi_i)^{N_f-1}(\phi_i\bar{\psi}^i)^{N_c-N_f}$: & $\nytab{k+\half N_f-N_c}{N_c+1-N_f}$ \\
& $N_f\le N_c < 2N_f-1$ & $w x^{k-N_f+N_c+1+kr}$ &  \\
\hline
1c & $(k>\half N_f, k+\half N_f>N_c);$ & 
$V_+\phi_1^{N_c^d+nN_f+m}(\lambda_1^i\phi_i)^{N_f-1}(\phi_i\bar{\psi}^i)^{N_f-1}$ 
& $\nysymtab{k+\half N_f-(N_c+1)} \times \nyantisymtab{m}$ \\
& $N_c+1=(n+1)N_f+m$, with & $(\del\phi_i\bar{\psi}^i)^{N_f}(\del^2\phi_i\bar{\psi}^i)^{N_f} \ldots (\del^{n-1}\phi_i\bar{\psi}^i)^{N_f}$ &  \\
& $n\geq 1,~~ 0\leq m<N_f$ & $(\del^n\phi_i\bar{\psi}^i)^{m}$: &  \\
&  & $w x^{k+N_f-N_c-1+N_f n(n+1)+2m(n+1)+kr}$ &  \\
\hline
\hline
2a & $(k\leq \half N_f, N_f \geq N_c);$   
& $V_+\phi_1 (\lambda_1^i \phi_i)^{k+\half N_f-1} 
(\phi_i\bar{\psi^i})^{N_c-(k+\half N_f)}$: & $\nytabbar{N_c+1-(k+\half N_f)}{N_f-N_c}$ \\
& $k+\half N_f < N_c$ & $w x^{-k+N_c+1+kr}$ &  \\
\hline
2b & $(k\le\half N_f, N_f \geq N_c);$   
& $V_+ \phi_1^{k+ {N_f \over 2}-N_c+1} (\lambda_1^i \phi_i)^{N_c-1} $: & $\nytab{k+\half N_f-N_c+1}{N_c}$ \\ 
& $k+\half N_f\ge N_c$ & $w x^{k+N_f-N_c+1+kr}$ &  \\
\hline
\end{tabular}
\caption{The leading ${\cal M}_+$ monopole operator in $U(N_c)_k(N_f,0)$ theories in various parameter ranges.}
\label{Chiralplus}
\end{table}

Depending on the sign of $k-{N_f \over 2}$ we have Case 1 and Case 2, which are further divided into subcases. Note that the results for Cases 1a and 2b are almost the same as the non-chiral Case 1 analyzed in the previous section, except for some shifts in the monopole charges. The typical lowest operator here consists of a bare monopole operator dressed by operators with the same charges as ($\phi_1^{k+{N_f \over 2}}$) so as to cancel the gauge charge. In other cases these operators do not survive, and the lowest one which does survive has more gauge-invariants attached to it (some appropriate number of $\phi_i \bar \psi^i$'s). We analytically analyze the contributions of the simplest monopole operators in section \ref{analyticone} below, while some of the other cases are analyzed in appendix \ref{apdx:AA}.


\subsection{Monopole GNO charge $(-1,0,\ldots)$ sector}

The charges of the bare monopole $V_-$ now correspond to
\begin{equation}
V_- \rightarrow z_1^{k - {N_f \over 2}}x^{-(N_c-1)+ (1-r){N_f \over 2}}.
\end{equation}
In Table \ref{Chiralminus} we give the results of Mathematica for the lowest monopole operator in each case. 
\begin{table}\centering
\begin{tabular}{| p{0.5cm}|  p{4.5cm} |c|c|c|}
\hline
&    Region   & Operator: $w^{\sum m_i} x^{\epsilon+j_3}$  & $SU(N_f)$ flavour rep.\\ \hline
1 a & \small{$k+{N_f\over 2}-N_c>0$} & $V_- {\bar \psi}^1(\lambda^1_i\bar{\psi}^i)^{k-\half N_f-1}(\phi_i \bar \psi^i)^{N_c-k+{N_f \over 2}}$: 
&  $\nytabbar{N_c+1}{k + {N_f \over 2}-N_c}$  \\ 
& $k > {N_f\over 2 }$  & $w^{-1}x^{k+N_c+1-k r}$ &  \\ 
& $N_f \ge N_c^d \equiv k+{N_f \over 2} - N_c$ &  &  \\ 
\hline
1 b & \small{$k+{N_f \over 2}-N_c>0$} & $V_-(\bar{\psi}^1)^{k-\half N_f-N_c+1}(\lambda^1_i\bar{\psi}^i)^{N_c-1}$: 
& $\nytabbar{N_c  }{k - {N_f \over 2} - N_c+1}$ \\ 
& $k > {N_f\over 2 }$ & $w^{-1}x^{3k+1-N_c-N_f-k r}$ &  \\
& $2N_f-1 >  N_c^d  \ge N_f$ &  &  \\
\hline
1 c & \small{$k+{N_f \over 2}-N_c>0$} 
& $V_-(\lambda^1_i\bpsi^i)^{N_c-1} (\bpsi^1)^{N_f}(\del\bpsi^1)^{N_f} \ldots$  
& $\nysymtabbar{N_c-1} \times \nyantisymtabbar{m}$  \\ 
& $k > {N_f\over 2 }$ & $~~~~~~~~~~~~~~~~~~~~~(\del^{n-1}\bpsi^1)^{N_f} 
(\del^n\bpsi^1)^m$: &  \\
& $ N_c^d+1=(n+1)N_f+ m$ & $w^{-1}x^{3k-N_c+1+N_f(n^2-n-1)+2mn-kr}$ &  \\
& where $n\geq 1$, ~$0\le m<N_f$  &  &  \\
\hline
\hline
2 a & ${ N_f  \over 2}- k > 0 ,  N_f > N_c$ 
& $V_- \phi_1^{{N_f\over 2}-k-N_c+1} (\lambda_1^i \phi_i )^{N_c-1}$: 
& $\nytab{{N_f \over 2}-k-N_c+1}{N_c}$ \\
& $k + {N_f \over 2} < N_f - N_c \equiv N_c^d$ & $w^{-1}x^{N_f - k+1-N_c- kr}$ &  \\ 
\hline
2 b & ${ N_f\over 2}-k > 0 ,  N_f > N_c$ 
& $V_-\phi_1(\lambda_1^i\phi_i)^{{ N_f\over 2}-k-1}(\phi_i \bar\psi^i)^{N_c-({N_f \over 2}-k)}$: 
& $\nytabbar{k-{N_f \over 2}+ N_c+1}{N_f-N_c}$ \\ 
& $k + {N_f \over 2} \ge  N_f - N_c \equiv N_c^d$  & $w^{-1}x^{k+ N_c+1-k r}$ &  \\
\hline
\end{tabular}
\caption{The leading ${\cal M}_-$ monopole operator in $U(N_c)_k(N_f,0)$ theories in various parameter ranges.}
\label{Chiralminus}
\end{table}
Again, depending on the sign of $k-{N_f \over 2}$, we have Case 1 and Case 2, which are further divided into subregimes.  The typical operator consists of a bare monopole operator dressed with fermions (bosons) if the sign of $k - {N_f \over 2}$  is positive (negative), so as to cancel the gauge charge of the bare monopole. Note that for $k=\frac{N_f}{2}$ the bare monopole $V_-$ is gauge-invariant by itself, and does not need to be dressed.
The details of the spins and R-charges of these operators are given in appendix \ref{apdx:DCM}. 
 
 
 \subsection{Consistency with duality}\label{consistencychiral}

 From our results above we can easily derive the result for $U(N_c)_{-k}(0,N_f)$ theories. Again, denoting the index of $U(N_c)$  CS theories with $N_f$ fundamental and $N_{a}$ antifundamental chiral multiplets by $I_{N_c,k,N_f,N_{a}}(r;x,w,t_a,\tilde t_a)$, one can see from \eqref{SCIdef1} that 
\begin{equation}
I_{N_c,k,N_f,0}(r;x,w,t_a,*)  =  I_{N_c,k,0,N_f}(r;x,w^{-1},*,  t_a)
= I_{N_c,-k,N_f,0}(r;x,w^{-1},t_a,*) .
\end{equation}

Now we can check that the results in the tables for the lowest surviving monopole operators with $\sum m_i = \pm 1$  are consistent with the expected dualities \cite{Benini:2011mf} (here these are exact dualities when $k \neq \frac{N_f}{2}$, not just for the lowest
monopole operators):
\begin{eqnarray}\label{ChiralMatch}
\nonumber
\text{Case }1: & k\geq \half N_f, \qquad
I_{N_c,k,N_f,0}(r;x,w,t_a,*) &=I_{N_c^d,-k,0,N_f}(1-r;x,{x^{-{N_f \over 2}} w^{-1}},* ,t^{-1}_a),  \\ 
\nonumber 
&(N_c^d=k+\half N_f-N_c)&=  I_{N_c^d,k,N_f,0}(1-r;x,{x^{-{N_f \over 2}} w^{-1}},t^{-1}_a,*), \\
\nonumber
\text{Case }2:&  k\leq \half N_f, \qquad
I_{N_c,k,N_f,0}(r;x,w,t_a,*) & = I_{N_c^d,-k,0,N_f}(1-r;x,{x^{-k} w^{-1}}, *,t^{-1}_a), \\ 
 &(N_c^d=N_f-N_c)&  =I_{N_c^d,k,N_f,0}(1-r;x,{x^{-k} w^{-1}}, t^{-1}_a,*).
\end{eqnarray}
The last equality implies that the duality should map the operators of Table \ref{Chiralplus} to the
ones of Table \ref{Chiralminus} for the same value of $k$, up to complex conjugation of the flavour representation and a shift in
the power of $x$. We find that
in these theories all the different subcases map to themselves under the duality.

The case $|k|=\half N_f$ is special as the bare monopole operator $V_-$
\footnote{Recall that we assume $k>0$. For $k<0$, $V_+$ is gauge-invariant.} is gauge-invariant in this case and hence survives in the $\sum m_i=-1$ 
sector as the lightest operator. Further, as in the non-chiral case with $k=0$, the duality matching 
works differently for $V_-$ as it maps (using the bottom lines of \eqref{ChiralMatch}) to an extra 
singlet chiral superfield $V_+$ in the dual theory, while the singlet
${\tilde V}_-$ is removed from the chiral spectrum of the dual theory by the superpotential $W = V_+ {\tilde V}_-$
\footnote{More precisely, the superpotential means that ${\tilde V}_-$ cancels in the index 
with a fermionic component ${\bar \psi}_{V_+}$ of $\bar{V_+}$, which sits in the same non-chiral multiplet.}.
The leading contribution in the $\sum_i m_i=1$ sector of the original theory is given 
by Case 1a (or 2b) of Table \ref{Chiralplus} for $k=\half N_f$. 
To find its dual one has to take into account the 
contribution of the gauge-singlet chiral multiplet $V_+$ in the dual theory. Using the results from 
a Mathematica computation, we claim that the dual of 
${\cal M}_+$ of the original theory actually  comes from the $\{m_i\}=0$ sector (recall that the
singlet $V_+$ also carries a $U(1)_J$ charge), and has the same
charges as
\be\label{cvpd}
{\cal M}_- : \quad  {\bar \psi}_{V_+} (\phi_i\bar{\psi}^i)^{N_c^d}.
\ee
Note that even though ${\bar \psi}_{V_+}$ is non-chiral as described above, this is not
necessarily true for its descendants or its products with other operators; for instance,
descendants by derivatives appear in the index for ${\bar \psi}_{V_+}$ but not for ${\tilde V}_-$,
and the latter operator can be separately multiplied by $(\phi_1 \bar{\psi}^1)$ which is a
singlet of $U(1)\times U(N_c^d-1)$, while the former operator in the $\{m_i \}=0$ sector cannot.


\section{The leading chiral monopole operators in the 't Hooft large $N_c$ limit}
\label{LNTL}

The ${\cal N}=2$ supersymmetric Chern-Simons-matter theories described above are particularly interesting
in the 't~Hooft large $N_c$ limit (keeping fixed $\lambda = N_c / k$ and $N_f$); in this
limit their thermal partition function can be computed exactly \cite{Aharony:2012ns}, and for finite
large $N_c$ we
can flow from the supersymmetric dualities to non-supersymmetric dualities
\cite{Jain:2013gza}.

For non-chiral theories in the 't Hooft limit,
the relevant case is Case 2 in Table  \ref{NonChiral}. Note that in this case
the leading monopole operator does not take the naively expected form, and includes
a large number of fermions. In this case, the scaling dimension of the lowest
chiral monopole operator scales as $N_c$ in the 't Hooft limit, as expected. 

For chiral theories in the 't Hooft limit, the relevant case is
Case 1c of Tables \ref{Chiralplus} and \ref{Chiralminus}. Also in this
case the monopole operators do not take the naive form, and include a large number of
fermions. Notice that 
for these theories the scaling dimension of the lowest chiral monopole operators 
scales as $N_c^2$ in the 't Hooft limit (since $n$ in the tables scales as $N_c$), unlike the non-chiral case. 
This implies that for this case all monopole operators with a dimension scaling as $N_c$ 
are not actually chiral. The difference between the two cases is that in the
non-chiral case we can use the operators $\tilde{\phi}$ to construct chiral
operators, but these are not available in our chiral case.

These conclusions can be avoided if we keep $(k-N_c)$ fixed in
the large $N_c$ limit (and in particular take $\lambda=1$). For instance, in the
non-chiral case if we take $k < N_c$ but also $N_c-k < N_f$ (as required to
preserve supersymmetry), then we are actually in Case 3 of Table \ref{NonChiral}.
The dual theory in this case has finite $N_c^d$ so it is not in the 't~Hooft limit.

\subsection{The mapping to high-spin gravity theories}

As we mentioned in the introduction, CS-matter theories are believed to be
dual to high-spin gravity theories, such that their 't~Hooft large $N_c$ limit
corresponds to classical high-spin gravity theories (see \cite{Giombi:2012ms}
for a review). States with high-spin gravity particles
correspond to operators with dimensions of order $1$ in the large
$N_c$ limit, while classical solutions of the high-spin gravity theories correspond
to operators with dimensions of order $N_c$ (recall that the coupling constants
in these theories are of order $1/N_c$). The same non-supersymmetric high-spin gravity theories
are dual to the CS-scalar and CS-fermion theories, and they have a
parameter $\theta_0$ that corresponds to the 't~Hooft coupling constant
of these theories (there is also a choice of boundary conditions that
determines whether the dual is a free theory coupled to CS, or a
critical one). The supersymmetric versions of these high-spin theories have
similar properties, and map to various supersymmetric CS-matter theories
(chiral or non-chiral) \cite{Chang:2012kt}.

The $U(1)_J$ global symmetry that the monopoles are charged under maps
on the gravity side to the $U(1)$ gauge field in the high-spin multiplet
(this multiplet, in the ``non-minimal'' high-spin theory, contains gauge
fields of all integer spins). This is true both in the supersymmetric and
in the non-supersymmetric cases. We thus expect classical solutions that
carry this charge to correspond to monopole operators with dimensions
of order $N_c$.

However, our arguments imply that such solutions should not exist in
many high-spin gravity theories. In the CS-fermion theories we argued
(see appendix \ref{fermion_counting}) that there are no monopole
operators with dimensions of order $N_c$ at large $N_c$, so no
such solutions should exist in the original non-supersymmetric high-spin
theory. In the ${\cal N}=2$ supersymmetric theories such monopole operators may
exist (and they certainly exist in the non-chiral theories), but we
argued that for the $N_a=0$ chiral theories all chiral monopole operators have
dimensions at least of order $N_c^2$. Thus, the corresponding
high-spin gravity theories should not have any classical BPS charged 
solutions. Note that even in the cases where monopoles do exist
with dimensions of order $N_c$, we expect these dimensions at
weak coupling to be at least of order $k = N_c / \lambda$, such
that they diverge in the $\lambda \to 0$ limit (which corresponds
to $\theta_0=0,\frac{\pi}{2}$; note that the coupling in the gravity
theory goes as $1/N_c$ rather than $1/k$ in this limit). Thus, in any high-spin gravity theory we
do not expect to have classical charged solutions in the parity-preserving
$\theta_0=0,\frac{\pi}{2}$ theories.
Note that we cannot say if specific monopole operators correspond to
classical gravity solutions or not, but when there is no monopole
operator there cannot be a corresponding gravity solution.

Some classical solutions of the non-supersymmetric high-spin gravity theories were
found in \cite{Didenko:2009td}, and were generalized to supersymmetric
cases (including some of our chiral and non-chiral theories) in 
\cite{Didenko:2009td,Bourdier:2014lya}. A linearized analysis suggests
that these solutions carry a charge under the $U(1)$ gauge symmetry
in the high-spin multiplet, but it is difficult to verify this. Our arguments
above imply that these solutions actually cannot carry this charge
(assuming that the duality to CS-matter theories is correct), and
it would be interesting to verify this directly.

Note that one way to avoid these arguments would be if the gravity
theories are actually dual to $SU(N_c)$ CS-matter theories, rather than
to $U(N_c)$ theories; as far as we know, none of the computations
performed up to now can distinguish between these two cases.
However, the $U(1)$ global symmetry has a very different interpretation
in the $SU(N_c)$ theories, where it is a baryon number symmetry (and
there is no global symmetry carried by monopoles). So, the arguments
above do not rule out classical charged solutions if the dual gauge theories
are $SU(N_c)$ theories. Naively, such theories should always have
baryons with dimensions of order $N_c$, which could correspond to
classical charged solutions on the gravity side. However, in the CS-scalar theories
this is actually not the case, because the baryon operator must be
anti-symmetric in the color index. An argument similar to the one in
appendix \ref{fermion_counting} then implies that when $N_f \ll N_c$ it must have a
dimension at least of order $N_c^{\frac{3}{2}}$. Thus, even if the gauge
group is $SU(N_c)$, we still claim that the non-supersymmetric
high-spin theories cannot have classical charged solutions (and in
particular this still means that the solutions of \cite{Didenko:2009td}
cannot be charged).

\section{Analytic arguments for chirality of $V_{+}\phi_1^k$ and related operators}
\label{analyticone}

The results presented in Tables \ref{NonChiral}, 
\ref{Chiralplus} and \ref{Chiralminus} 
for the lowest lying monopole operators for non-chiral and chiral theories, 
respectively, are conjectural and based on extrapolating Mathematica computations done for low values 
of $N_c$, $k$ and $N_f$. In this section we present analytic arguments 
for the simplest operators of the schematic form $V_{+}\phi_1^k$
\footnote{Here we assume $k>0$. For a more detailed discussion, and for similar 
arguments for some other operators, see appendix \ref{apdx:AA}.}. 

Let us consider the GNO sector $\{m_i\}=\{+1,\orao\}$ in the $U(N_c)_k(N_f,N_f)$ theory. 
The relevant supersymmetric letters to build gauge-invariant operators with the same $(\E+j_3)$ 
and axial charge as $V_{+}\phi_1^k$ (in this monopole background) have $(\epsilon,j_3,R)$
equal to:
\be\label{nc1a}
\phi_1 \ra \left( \half +r,\half,r \right), \quad \phi_i \ra ( r,0,r ), \quad
\lambda_1^i \ra (1,0,1).
\ee
Using these letters we want to construct $U(1)\times U(N_c-1)$ gauge-invariant operators.

Notice that replacing any of the $\phi_1$'s with $\lambda_1^i\phi_i$ keeps the $x$ 
and $y$ charges. Since there are only $(N_c-1)$ $\lambda_1^i$'s, and they are anti-commuting,
the maximum number of $\phi_1$'s that
one can replace with $\lambda_1^i\phi_i$ is $\min(N_c-1,k)$. It is easy 
convince oneself that the operators generated in this way 
exhaust all the naively chiral operators at this level (this power of $x$). Furthermore, each such 
replacement flips the sign of $(-1)^F$ and also changes the $SU(N_f)_l$ flavour 
representation, since the $\phi$'s are symmetric in flavour, while the $(\lambda \phi)$'s are
anti-symmetric. Thus there are potential cancellations,
and whether or not any contribution survives at this level depends on whether all these 
flavour representations cancel or not. Whenever there is a cancellation we expect that the
corresponding bosonic and fermionic operators (that have the same global charges) join
together into a single non-chiral multiplet of the superconformal algebra. In this section we perform this analysis.

The total Index contribution of all the operators at this level 
can be schematically written as 
\be
\sum_{n=0}^{\min(N_c-1,k)} (-1)^n (\phi_1^a)^{k-n}(\lambda_1^i\phi_i^a)^n.
\ee
Note that for $n > N_f$ these operators vanish due to anti-symmetry of the last
factor in $SU(N_f)_l$.
Since the $x$ and $y$ charges of all these operators are the same, looking only at the 
$SU(N_f)_l$ flavour representations (these operators are singlets of $SU(N_f)_r$), we get
\be\begin{split}\label{ncc2rep}
= & \sum_{n=0}^{\min(N_c-1,k,N_f)} (-1)^n\left( \quad \nysymtab{k-n} \otimes \nyantisymtab{n} \quad \right) \\
= & \sum_{n=0}^{\min(N_c-1,k,N_f)} (-1)^n\left( \nytab{k-n}{n+1} \oplus \nytab{k-n+1}{n} \right) \\
= & \left\{
\pbox{90mm}{
$(-1)^{N_c-1} \nytab{k-N_c+1}{N_c}  \quad \text{if} \quad k\geq N_c \text{ and } N_f\geq N_c$ \\
\\
$0 \hspace{45mm} \text{if} \quad k<N_c \text{ or } N_f<N_c$ \\
} \right\}.
\end{split}\ee 
In the first line the $(k-n)$-box symmetric representation comes from $(\phi_1^a)^{k-n}$, while 
the $n$-box antisymmetric representation comes from $(\lambda_1^i\phi_i^a)^n$ (taking
into account the anticommutation of the $\lambda_1^i$'s).
The second line gives the decomposition into irreducible representations of the tensor 
product in the first line. The third line uses the fact that representations cancel 
pairwise between the $n$'th and $(n+1)$'th terms, and the only (if at all) non-canceling 
contribution comes from the last term in the series when $n=N_c-1$. This is precisely Case 1 of Table \ref{NonChiral}.


For two of the remaining three cases, namely Cases $2$ and $4$ listed in 
Table \ref{NonChiral}, we will present similar but 
slightly more involved analytic arguments in appendix \ref{apdx:AA}.

The argument presented above can be straightforwardly applied for similar 
operators in chiral theories as well. In a $U(N_c)_k(N_f,0)$ theory the 
corresponding operators are $V_{+}\phi_1^{k+\half N_f}$, as the gauge 
charge of the bare monopole operator $V_{+}$ is $-(k+\half N_f)$. The only 
difference here is thus a shift of $k$ by $\half N_f$. The above argument then 
implies that a non-vanishing contribution at this level occurs for 
\be\label{mtncex}
k+\half N_f \geq N_c \quad \text{and} \quad N_f\geq N_c,
\ee
and the surviving $SU(N_f)$ representation is 
\be
 \nytab{k+\half N_f-N_c+1}{N_c}.
\ee
This gives Cases 1a and 2b in Table \ref{Chiralplus}.
Notice that the conditions in \eqref{mtncex} imply that the rank of the 
dual gauge theory is non-negative, which is required for unbroken supersymmetry.

In appendix \ref{sec:ncaa} we present similar arguments for a subset of the other 
cases listed in Tables \ref{Chiralplus} and \ref{Chiralminus}.

\section{A possible dual of $V_{+}\phi_1^k$ when it is not chiral}
\label{nonchiraldual}

In this section we discuss how the dual operator to $V_+ \phi_1^k$ looks like in
non-chiral $U(N_c)_k (N_f,N_f)$ theories, when
this operator is not chiral (which is true for all $N_c > 1$).
Since the dualities in Chern-Simons-matter theories are strong-weak dualities, in the case
where the operator $V_+ \phi_1^k$ is not chiral,
the operator dual to it will in general have a very different weak coupling 
scaling dimension. But it must have the same 
values of the other global charges, namely spin, axial charge and flavour representation. 
Moreover, since we expect $V_+ \phi_1^k$ to be the lowest operator with the same
quantum numbers even when it is not chiral, we expect it to be dual to the lowest
operator with these quantum numbers in the dual theory, because there should be
no level-crossing of the operators in a fixed representation\footnote{This argument is
not rigorous, because in CS-matter theories the coupling constant that we use to
go from weak to strong coupling is discrete, rather than continuous. However, we do expect it
to be valid at least in the 't Hooft large $N_c$ limit, where this parameter becomes
effectively continuous.}.

For $N_c > 1$, the operator $V_+ \phi_1^k$ sits in the $k$-box
symmetric $SU(N_f)_l$ flavour representation and is not chiral. To find its dual we need to look for operators in the $U(N_c^d)_k$
theory which have
\begin{itemize}
\item $j_3=\half k$,
\item A conjugate symmetric $k$-box representation under the $SU(N_f)$ flavour symmetry acting
on the ${\tilde \phi}$'s of the dual theory,
\item Axial charge = $k-N_f$,
\item R-charge = $N_f - N_c + 1 + r(k-N_f)$,
\end{itemize}
where the axial charge and R-charge are those of the original theory.
Note that when the operator $V_{+}\phi_1^k$ is not chiral, there is no reason 
for the dual operator to be constructed out of only the supersymmetric letters
that we discussed until now. 
Allowing for non-supersymmetric letters of the dual theory (taken to have level $k$ as in the second line of \eqref{nonchiralduality})  the simplest 
possible operator with the same axial charge, R-charge, $j_3$ and flavour representation is 
\be \label{nonChiral}
V_-^\dagger (\tilde {\bar\phi}_1)^k  \lambda_1^i \lambda_i^1.
\ee 
Since $V_-^\dagger$ and $ \lambda_1^i \lambda_i^1$ have $j_3=0$ and are flavour singlets, the spin and flavour 
representations match trivially (if we choose the $\tilde {\bar\phi}_1$'s in the monopole background
to have $j_3=\frac{1}{2}$). Note that the gauge charge of $V_-^\dagger$ is $-k$. The $ \lambda_1^i\lambda_i^1$ factor is just to compensate for the 
R-charge. So, we conjecture that this operator is dual to $V_+ \phi_1^k$ in the original
theory.

In the chiral $U(N_c)_{N_f,0}$ case, the above argument goes through except for the minor change that 
here the original theory has the operator $V_+ \phi_1^{k+{N_f \over 2}}$. In the dual theory (taken
to have level $k$ as in \eqref{ChiralMatch}), 
we look for operators with topological charge $\sum m_i = -1$. The only subtlety is that 
the R-charge of the dual theory is shifted by $N_f \over 2$ as in 
\eqref{ChiralMatch} compared to the original theory. Keeping track of this 
shift, the obvious candidate for the dual operator is 
\be
V_+^\dagger ({\bar\phi}_1)^{k + {N_f \over 2}}  \lambda_1^i \lambda_i^1.
\ee
Note that the gauge charge of $V_+^\dagger$ is $k + {N_f \over 2}$. The spin matches if we again 
choose the ${\bar\phi}_1$'s in the monopole background to have $j_3=\frac{1}{2}$.

\section{Perturbative corrections to $V_{+}\phi_1^k$ in Chern-Simons-scalar theories}
\label{nspc}

In this section, we return to our original motivation of understanding the mismatch of the $N_c$ scaling
of the classical dimensions of monopole operators under the non-supersymmetric Chern-Simons duality in the 't~Hooft large $N_c$ limit.  

Consider the monopole operator $V_+\phi_1^k$ in a $U(N_c)_k$ Chern-Simons theory coupled to a single scalar field (the analysis is similar
for theories with fermions, except that there already the classical dimension scales as $|k|^{\frac{3}{2}}$ for large $|k|$). 
Using radial quantization, the scaling dimension of any local operator in the flat space theory 
is mapped to the energy of the corresponding state on $S^2$. The operator $V_+\phi_1^k$  corresponds to a state with unit magnetic flux on $S^2$, with $k$ lowest energy scalar $\phi_1$ modes excited to neutralize the charge of the bare flux state. 

For operators of this type, whose classical energy scales as $N_c$ in the 't Hooft large $N_c$ limit (in which $\lambda \equiv N_c/k$ is kept fixed),
one expects perturbation theory not to be valid, and perturbative corrections to the energy to also be of order $N_c$ (see, for instance,
\cite{Kim:2009ia}; this is the case even when classical solutions for these monopoles exist in the Chern-Simons-matter theory,
as in \cite{Kim:2009ia,Kim:2010ac} \footnote{The appendix of \cite{Kim:2010ac} constructs classical BPS monopole solutions for
the ${\cal N}=2$ supersymmetric theories we discuss in this paper, that correspond to operators like $V_+ \Phi^k$. As discussed above, in most cases we
expect this operator not to be chiral in the full theory, and then the corresponding classical solutions could also acquire
large quantum corrections.}). The general arguments are very similar to the analysis of baryons in the large $N_c$ limit of QCD \cite{Witten:1979kh}, and
we will discuss this analogy further below. However, at least in some cases one expects such operators to correspond to
classical solutions of some `master field' theory whose coupling constants scale as $1/N_c$ (an example of this is the Skyrme
model description of baryons in QCD; in our theories the role of this `master field' theory is played by the dual high-spin
gravity theory). In this context one may expect corrections to the dimensions coming
from the classical solutions (which are of order $N_c$) to be suppressed by powers of $1/N_c$, such
that the energy of these configurations would remain of order $N_c$ in the `t~Hooft large $N_c$ limit. In our case, as we
discussed, such a scaling does not seem to be consistent with duality. In this section we will argue that the perturbative corrections 
to the anomalous dimensions of monopole operators might violate the naive large $N_c$ counting, even at very weak coupling (this
implies that these operators do not correspond to classical solutions of any `master field' theory). 

\begin{figure}[h]
\centering
\begin{tabular}{cc}
\includegraphics[width=70mm,height=40mm]{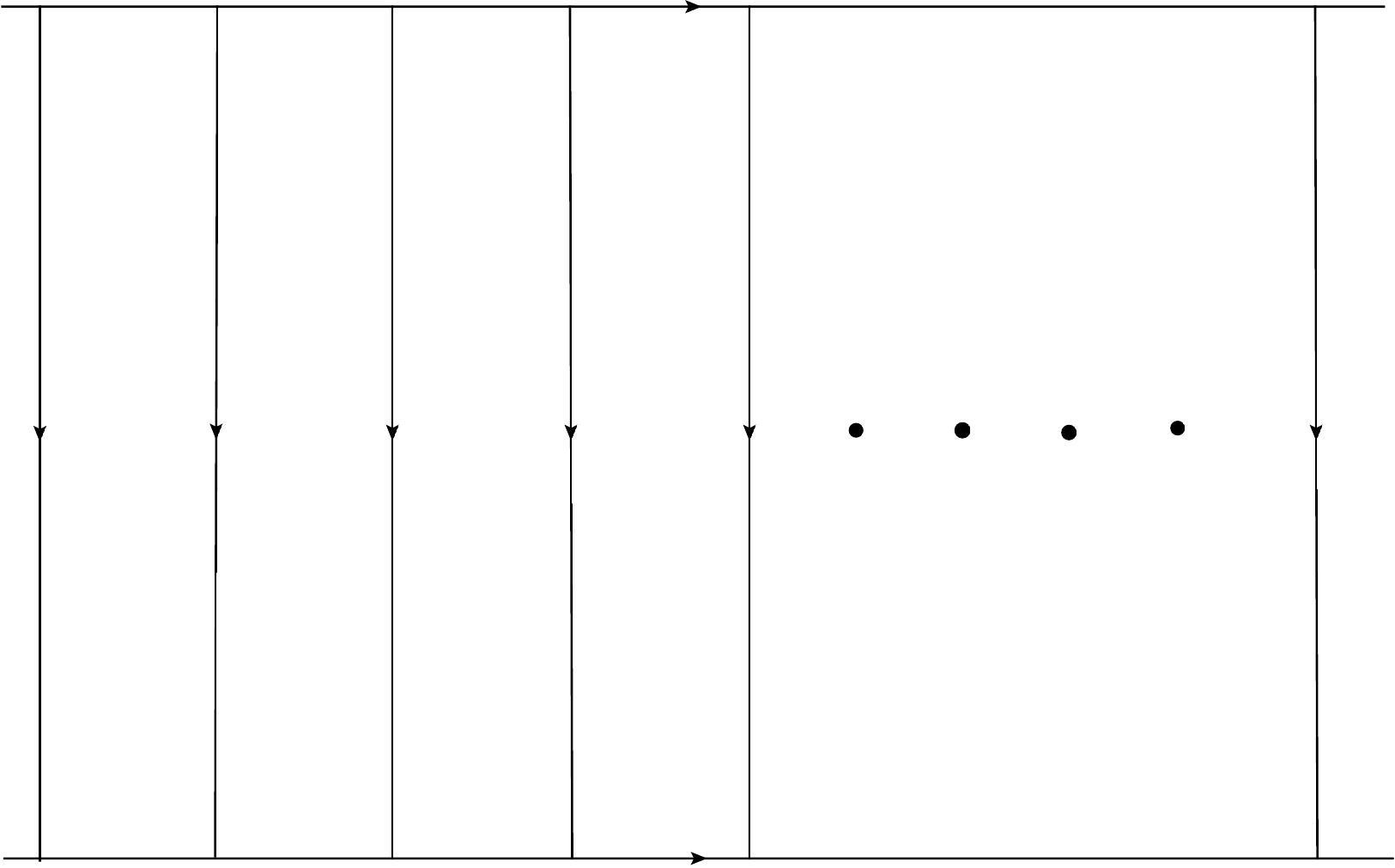} & \includegraphics[width=70mm,height=40mm]{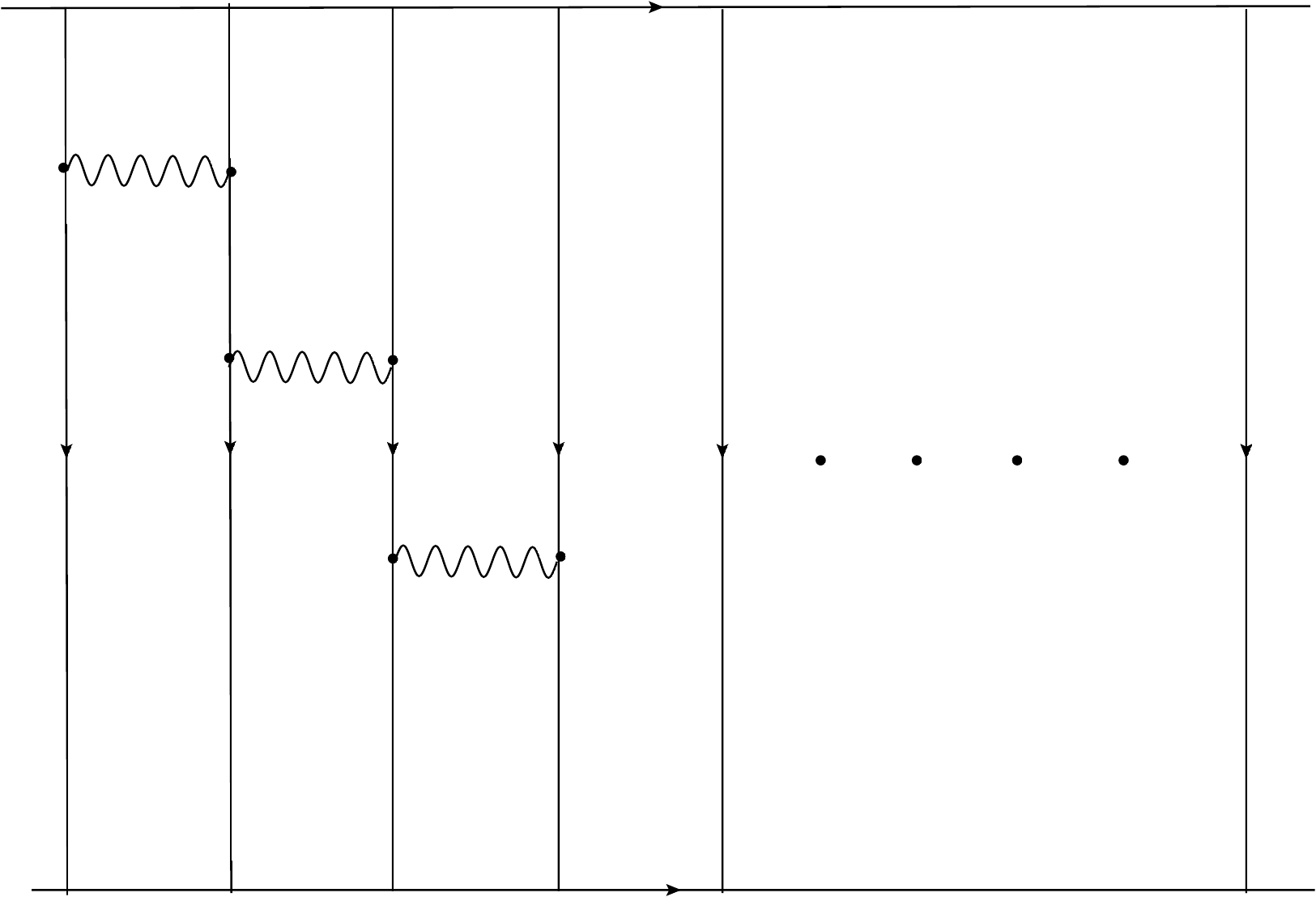} \\
a & b \\
\includegraphics[width=70mm,height=40mm]{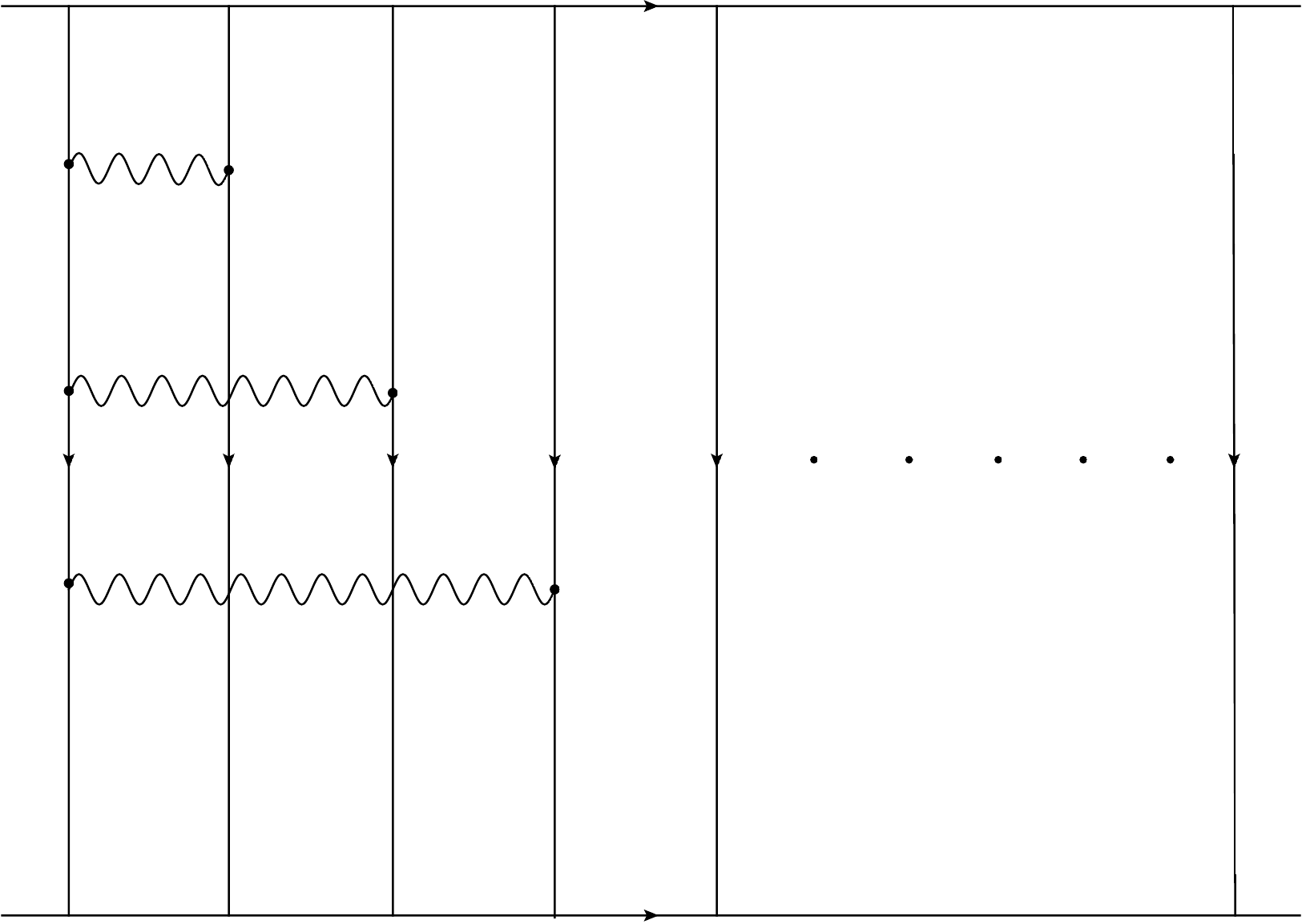} & \includegraphics[width=70mm,height=40mm]{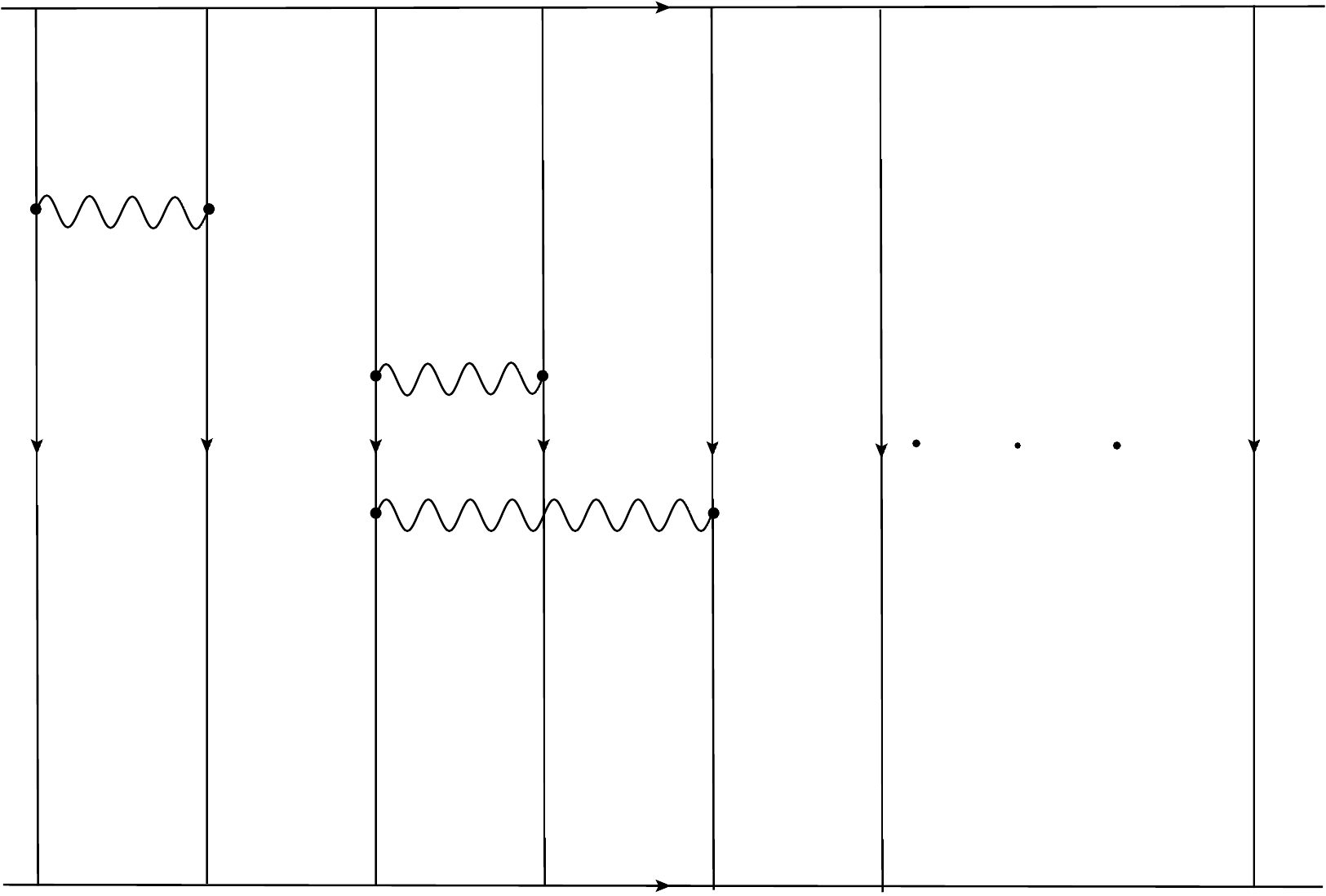} \\
c & d \\
\end{tabular}
\caption{Some of the diagrams contributing to the energy of the lowest flux state, which is related to the dimension of the leading monopole operator.}
\label{fig:pfigs}
\end{figure}

We will work with a normalization of the action where the gauge propagators come with $k^{-1}$, while 
the scalar propagators and interaction vertices have no factors of $k$ or $N_c$ (the 3-gluon vertex
scales as $k$). In the 't Hooft large 
$N_c$ limit at weak coupling ($\lambda \rightarrow 0$) we can restrict to planar diagrams with no loops 
(each loop comes with a factor of $\lambda$ and hence is suppressed at weak coupling). Thus, the leading contribution 
to the ground state energy of a unit flux state comes from the diagrams of the form shown in figure 
\ref{fig:pfigs}(a,b,c) with an arbitrary number of horizontal gluon propagators \footnote{The Chern-Simons-scalar theories
also have $(|\phi|^2)^3$ couplings whose coefficients scale as $1/k^2$ in the small $\lambda$ limit \cite{Aharony:2011jz}. Thus, there are
also diagrams with vertices of this type and with no loops, which contribute at the same order as the diagrams with gluons, and do
not modify our qualitative arguments.}.  The $k$ vertical lines here are scalar propagators,
which all have the same color index. We will show below that all these diagrams have a contribution to the energy that has the same scaling with $k$ at large $k$; note that the connections between different scalar lines do not have to be planar (as figure \ref{fig:pfigs}(c) illustrates). Furthermore, the diagrams can 
be divided into two subclasses: ``connected" and ``disconnected". The diagrams which are ``connected'' 
have the property that one can reach any of the gluons from any other gluon by only moving along  
the vertical scalar propagators and the horizontal gluon propagators, without having to go through 
the horizontal lines at the top or bottom. All other diagrams are ``disconnected''. When computing the evolution of the
monopole state for a time $T$, the ``connected'' diagrams scale as $T$ (compared to figure \ref{fig:pfigs}(a)), while all other diagrams scale
as higher powers of $T$. Thus, diagrams which 
are ``connected" in the above sense contribute to the energy of 
the state directly, while those which are ``disconnected" result from the expansion of the exponential 
of ``connected'' diagrams (they give $e^{i H T}$ in the partition function with a time-difference $T$). Some leading ``connected" diagrams are shown in figures 
\ref{fig:pfigs}(a,b,c) while figure \ref{fig:pfigs}(d) is an example of a ``disconnected'' 
diagram.

Even within the restricted class of ``connected'' diagrams, the number of diagrams with a 
given number of gluon propagators grows very fast \footnote{One can 
estimate that for $n$ gluon propagators, the number of relevant diagrams is related to 
the number of integer partitions of $n$, which grows exponentially for large $n$.}.  
We will not compute these diagrams explicitly but just perform an estimate of a 
subset of these diagrams to show that quantum corrections can potentially change 
the large $k$ scaling of the dimension of $V_+\phi_1^k$ (we assume here
that there is no reason for these corrections to all cancel, as would be the case
for chiral monopole operators in supersymmetric theories).

Let us normalize the contribution of the leading 
diagram without gluon propagators, figure \ref{fig:pfigs}(a), to be $k!$ (this 
is just the number of ways of contracting $k$ $ \phi$'s with $k$ $ \bar{\phi}$'s). Let 
us also restrict to the subset of diagrams of the type shown in figure \ref{fig:pfigs}(b). 
Such a diagram with $n$ gluons comes with a combinatoric factor of 
\be\label{ngdc}
\frac{(k(k-1)\cdots(k-n))^2}{2k^n}(k-n-1)! = k(k!)\prod_{l=1}^{n}\left(1-\frac{l}{k}\right).
\ee
Each factor of $k(k-1)\cdots (k-n)$ comes from the need to choose which scalar (anti-scalar)
connects to the first gluon line, which to the second gluon line, and so on (and we get a
factor of $\frac{1}{2}$ by inverting the order of the gluon lines). The factor of $k^n$ in the
denominator comes from the gluon propagators, and the factor of $(k-n-1)!$ comes
from the possible contractions of all the scalars that are not attached to gluon lines. Thus,
dividing by the diagram of figure \ref{fig:pfigs}(a), any such diagram with $n \ll \sqrt{k}$ gives a contribution
of order $k$ to the energy, which is the expected scaling of the monopole dimension. Note
that this has no powers of $\lambda$, so these diagrams contribute even at very
weak coupling (namely, in the limit of large $k$ with finite $N_c$).

For the purpose of our estimation we assume that the full contribution from such a
diagram  differs from the above combinatoric factor by an ${\cal O}(1)$ number, since
there are no obvious large factors involved. The contribution to the monopole energy from 
these diagrams is then estimated by summing over the contributions of 
this subset of ``connected'' diagrams. We expect that the approximation of the sum over 
``disconnected'' diagrams by the exponential of the ``connected'' diagrams should be good 
at least for ``connected'' 
diagrams with up to $\sqrt{k}$ gluon propagators. Assuming that all extra factors
are equal for all these diagrams, they sum up to
\be\label{sumest}
\sum_{n=1}^{\sqrt{k}} k(k!)\prod_{l=1}^{n}(1-\frac{l}{k}) \sim k^{\thalf} (k!).
\ee
It is easy to verify numerically that correcting this sum by a similar contribution from 
diagrams with a higher number of gluon propagators (by including ``connected'' 
diagrams in \eqref{sumest} with a number of gluon propagators larger then $\sqrt{k}$) does not
affect the leading large $k$ behaviour of the sum.

As described above, this is just an estimate for a very small subset of the leading diagrams at
large $k$. Taking into account all of the other leading diagrams could generate an even larger 
change in the large $k$ scaling of the monopole ground state energy compared to the ``classical" value.
On the other hand, clearly there is no reason to expect all these diagrams to be equal (or
even to have the same sign) as we assumed. But anyway, 
this shows that quantum corrections could affect the scaling dimensions of these monopole operators 
in a very drastic way in the large $k$ limit. In particular, we see that quantum corrections can potentially lead to 
a change in the $k$ scaling of the dimension of monopole operators, which could resolve the puzzle stated in the introduction regarding the difference in the large
$k$ (large $N_c$) scaling of the naive dimensions of the leading monopole operators across the dual pair of 
bosonic and fermionic Chern-Simons theories.

Note that the analysis above is very similar to the analysis of the masses of baryons in large $N_c$ QCD \cite{Witten:1979kh}. 
At weak coupling and 
leading order in large $N_c$ the masses of baryons are ${\cal O}(N_c)$. All the 
diagrams shown above will also contribute to the masses of baryons (note that in the baryon
case the quarks that replace the scalars all have different color indices, but because they are
contracted with an epsilon symbol, their wave function is eventually symmetric, just like the
one of the scalars in our case). Though 
the same diagrams are suppressed by a factor of $\lambda^n$, where $\lambda$ is the 't Hooft coupling\footnote{In large $N_c$ QCD the gluon propagator goes as $g_{YM}^2$ rather than as $1/N_c$, while the combinatorics is the same as above.},
they are all 
comparable for a coupling of ${\cal O}(1)$, and the above argument would suggest that 
the masses of baryons could change from ${\cal O}(N_c)$ to some higher 
power  of $N_c$ (at least when their spin is of order $N_c$). It is widely believed that this is not the case for baryons in the large $N_c$ limit 
\cite{Witten:1979kh}, and there is substantial evidence for this. However, there are various differences between our case
and that of baryons that could lead to a difference in the scaling of the two cases.


\section{Summary}
\label{summary}

In this paper we discussed monopole operators in Chern-Simons-matter theories. We discussed
in detail the chiral monopole operators in such theories with ${\cal N}=2$ supersymmetry, and
showed that in many cases the lowest monopole operator is rather different from the simplest
naively-chiral monopole operator. In the non-supersymmetric case we noted that duality implies
large corrections to the dimensions of monopole operators, which are naively inconsistent with
the 't Hooft large $N_c$ limit. We argued that these operators could have large perturbative
corrections to their dimensions, that may invalidate the usual large $N_c$ counting. It would be interesting to make these
arguments more precise, and perhaps even to compute the monopole dimensions for large $N_c$
and to verify that they are consistent with non-supersymmetric dualities.

There are many possible generalizations of our analysis. We analyzed only theories with $N_a=N_f$
or $N_a=0$, and the generalization to arbitrary values of $N_a$ and $N_f$ should be
straightforward. We also discussed only the simplest monopole operators with $U(1)_J$ charge
$\sum m_i = \pm 1$, and it would be interesting to generalize our analysis to higher charges.
It would be interesting to study the chiral rings in these theories including the monopole operators,
as done for some theories in \cite{Bashkirov:2011vy}
\footnote{
	Naively, one may expect to find a moduli space whenever there is a scalar chiral
	operator (appearing in the chiral ring). In the case of monopole operators,
	the corresponding branch of the moduli space would include a vacuum
	expectation value for the scalar in the vector multiplet, but we know that in the
	theories we discuss with $k \neq 0$ there is no such moduli space. Hence, we expect that even if
	some of our theories have scalar chiral monopole operators,
	the chiral ring would not be freely generated
	by these operators, but rather they have to vanish when raised to some power.
	In our analysis of the lightest monopole operators, we only found a scalar operator
	in Case 4, and in this case it is easy to believe that this operator raised to some power would
	vanish because fermions are explicilty involved in its construction.}.
It is not clear if all chiral monopoles with
$\sum m_i=1$ are products of the leading chiral monopoles we found with operators in the
$\sum m_i=0$ sector, and if all chiral monopoles with $\sum m_i > 1$ can be written as products
of chiral monopoles with $\sum m_i=1$.
One can also use similar methods to study theories with product gauge groups. Theories
of this type with a larger amount of supersymmetry  were analyzed in detail in the
literature, but most of the discussion in the literature (except \cite{Kim:2010ac})
is about monopoles that have rather different properties from the monopoles we discuss here.

For gauge groups that do not involve $U(N_c)$, it is not obvious how to identify the monopole
operators, since there is no $U(1)_J$ symmetry. Nevertheless, the index in these theories
is still written as a sum over monopole sectors with different GNO charges, and it would be
interesting to try to possibly identify and match different monopole states also for such other gauge groups.
In particular it would be interesting to do this for $SU(N_c)$ gauge theories, noting that their
analysis is completely different from the $U(N_c)$ analysis we presented here. In these theories
there is no $U(1)_J$ symmetry, but there is a $U(1)_B$ baryon-number symmetry whose gauging
leads to the $U(N_c)$ theory, and it would be interesting to use the index to understand which
baryon operators are chiral.

\section*{Acknowledgements}
We would like to thank Lorenzo Di Pietro, Nadav Drukker,  Zohar Komargodski, Silviu Pufu, Cobi Sonnenschein and Shimon Yankielowicz for useful discussions, and especially Guy Gur-Ari, Shiraz Minwalla and Ran Yacoby for useful discussions and comments on a draft of this paper.
This work was supported in part by an Israel Science Foundation center for excellence grant (grant no. 1989/14), by the Minerva foundation with funding from the Federal German Ministry for Education and Research, by the I-CORE program of the Planning and Budgeting Committee and the Israel Science Foundation (grant number 1937/12), by the Henry Gutwirth Fund for Research, and by the ISF within the ISF-UGC joint research program framework (grant no. 1200/14). OA is the incumbent of the Samuel Sebba Professorial Chair of Pure and Applied Physics.


\appendix

\section{Analytic arguments for chirality of monopole operators}\label{apdx:AA}

In this section we present analytic arguments for the leading chiral monopole operators in the $(\pm 1,\orao)$ sector,
for the cases when the leading operator does not involve derivatives or gauge-invariants of the form ($\phi_i \bar\psi^i$) 
attached to gauge-invariant monopole operators.

The monopole operators with the background flux $(1,\orao)$ break the $U(N_c)$ gauge symmetry 
to $U(1)\times U(N_c-1)$. In this background we have (in the weakly coupled high-energy
Yang-Mills-Chern-Simons theory which is used to compute the Index) $2(N_c-1)$ lightest gaugino states coming from
$\lambda_1^i, \lambda_i^1$, with $j_3=0$, transforming under the unbroken $U(1)\otimes U(N_c-1)$ as
\be\label{gzmc}
\lambda_1^i : \quad (-1, N_c-1), \qquad
\lambda_i^1 : \quad (+1, \overline{N_c-1}).
\ee
The contribution of these operators corresponds to the following factor in the Index:
\be
\prod_{i=2}^{N_c}\left(1-\frac{z_1}{z_i}x\right)\left(1-\frac{z_i}{z_1}x\right).
\ee
The other similar factors from the product over $i\neq j > 1$ constitute the Haar measure for the unbroken $U(N_c-1)$. Thus, in the
$(1,\orao)$ 
sector, the gauge-invariant operators are constructed as $U(1)\otimes U(N_c-1)$-invariants with 
$\lambda_1^i$ and $\lambda_i^1$ as additional 
supersymmetric letters (compared to the zero flux sector).

\subsection{$U(N_c)_k(N_f,N_f)$ theories}
Our strategy to construct gauge-invariant chiral monopole operators will be exactly the 
same as the usual construction of local gauge-invariant chiral non-monopole operators 
in perturbative gauge theories, i.e. to first identify the basic supersymmetric 
letters that can contribute to the index\footnote{Including the ``bare" monopole operator, transforming 
under some gauge group representation.}, and then to form combinations of these letters that give 
$U(1)\times U(N_c-1)$-invariant operators. 

To proceed, let us first identify the basic supersymmetric letters in the $(1,\orao)$ sector, satisfying 
$\epsilon=j_3+R$. Using \eqref{letters}, their  $(\epsilon,j_3,R)$ values are:
\be\begin{split}\label{bslac}
\phi_1,~\tilde{\phi}^1 & \ra \left(\half+r,\half,r\right), \\  
\phi_i,~\tilde{\phi}^i  & \ra \left(r,0,r\right), \\
\bar{\psi}_+^1,~\tilde{\bar{\psi}}_+^1 & \ra (2-r,1,1-r), \\
\bar{\psi}_+^i,~\tilde{\bar{\psi}}_+^i & \ra \left(\thalf-r,\half,1-r\right), \\ 
(\lambda_-)_1^i,~(\lambda_-)_i^1 & \ra (1,0,1), \\
\del_{++} & \ra (1,1,0). 
\end{split}\ee
Since the bare monopole operator is charged only under the $U(1)$ and is invariant 
under $U(N_c-1)$, combinations of basic letters which are $U(N_c-1)$ invariant but carry 
$U(1)$ charges are relevant for our counting purpose. Apart from these we also have
combinations of letters invariant under the full 
$U(1)\times U(N_c-1)$ gauge group which we need to use. 
All these combinations and their charges and flavour representations 
relevant for the Index are listed in Table \eqref{ncblc}.
\begin{table}
\centering
\begin{tabular}{|c|c|c|c|c|}
\hline
Operator & $U(1)$ & $\epsilon +j_3$ & $A$ & $(SU(N_f)_l, SU(N_f)_r)$ \\
\hline
$V_{1,\orao}$ & $-k$ & $(1-r)N_f-N_c+1$ & $-N_f$ & $({\cal I},{\cal I})$ \\ 
\hline
$\phi_1,~\lambda_1^i\phi_i$ & $+1$ & $1+r$ & $1$ & $(N_f,{\cal I})$ \\
\hline
$\tphi^1,~\lambda_i^1\tphi^i$ & $-1$ & $1+r$ & $1$ & $({\cal I},N_f)$ \\
\hline
$\bpsi^1,~\lambda_i^1\bpsi^i$ & $-1$ & $3-r$ & $-1$ & $(\bar{N_f},{\cal I})$ \\
\hline
$\tilde{\bpsi}^1,~\lambda_i^1\tilde{\bpsi}^i$ & $+1$ & $3-r$ & $-1$ & $({\cal I},\bar{N_f})$ \\
\hline
$\del_{++} $ & $0$ & $2$ & $0$ & $({\cal I},{\cal I})$ \\
\hline
$\phi_i\bar{\psi}^i$ & $0$ & $2$ & $0$ & $({\cal I}\oplus adj,{\cal I})$ \\
\hline
$\tphi^i \bar{\tpsi}_i$ & $0$ & $2$ & $0$ & $({\cal I},{\cal I}\oplus adj)$ \\
\hline
\end{tabular}
\label{ncblc}
\caption{Some of the basic $U(N_c-1)$-invariant combinations of supersymmetric letters and their gauge and global charges. Summations over the index $i$ run from $2$ to $N_c$.}
\end{table}

In this subsection we concentrate on monopole operators with $\sum_i m_i=+1$. 
This is due to the invariance of the Index of these non-chiral 
theories under
\be
\{m_i\} \ra \{-m_i\}, \quad \{z_i\} \ra \{z_i^{-1}\}, \quad \{w \ra w^{-1}\}, \quad \{t_a \leftrightarrow {\tilde t}_a\}.
\ee
This just says that for each operator in a given $\{m_i\}$ sector one can obtain a monopole
operator with $\{-m_i\}$  by interchanging
\be
(\phi,\bar\psi) \lra (\tphi,\tilde{\bar{\psi}}).
\ee

Another useful property of the Index is its invariance under a second set 
of transformations,
\be
k \ra -k, \quad \{m_i\} \ra \{-m_i\}, \quad \{w \ra w^{-1}\}.
\ee
Using this property we can restrict to $k>0$. 

Yet another useful property of the Index of these theories is the fact that the R-charge 
in these theories can be shifted by mixing it with the $U(1)_A$. Specifically, 
$y\ra y x^{r_0}$ shifts the R-charge by $r\ra r+r_0$. This freedom can be used to 
set the R-charge $r$ of $\phi,{\tilde \phi}$ to any convenient value, but we will not use this here.

\subsubsection{Case 1: $k\geq N_c$ and $N_f\geq N_c$}\label{sec:ncc1}

This case has already been discussed in section \ref{analyticone}. The surviving $SU(N_c)_l$ flavour 
representation in this case is 
\be\begin{split}\label{ncc1rep}
(-1)^{N_c-1} \nytab{k-N_c+1}{N_c}.
\end{split}\ee 
This representation survives only when 
\be\label{nc1c}
k\geq N_c \quad \text{and} \quad  N_f \geq N_c.
\ee
If any of these two conditions is violated, cancellation 
at this level is complete and we have to look for other  lightest operators.

In the next two subsubsections we give arguments for the lightest monopole
operators in the non-chiral theories for Cases 4 and 2, respectively.

\subsubsection{Case 4: $m\equiv N_c-k>0$ and $N_f\geq N_c$}\label{sec:ncc2}

In this case we will show that the leading monopole operator appears
at the level of $V_+\phi_1^k(\tphi^i\phi_i)^m$. The Index contribution 
of all the operators of this general form can be schematically arranged as the 
following series:
\be\begin{split}
\sum_n (\tphi^i\phi_i)^n \left(\sum_{l=0}^{k} (-1)^l \phi_1^{k-l} 
                    (\lambda_1^i\phi_i)^l \right).
\end{split}\ee
Naively from the above series it looks like that the Index contribution at 
this level also vanishes, as the series in the bracket vanishes for $k<N_c$, using the 
arguments given earlier for Case 1 in section \ref{analyticone}. But notice that to 
actually evaluate the Index contribution of this series we have to take the tensor product 
of the flavour representations of each term in the series with that of the 
$\phi_i^n$, taking into account that we only have $(N_c-1)$ different 
$\phi_i$'s, and hence more than this number cannot be antisymmetrized 
in constructing the flavour representations. This makes a difference
only when the total number of $\phi_i$'s in the 
operator exceeds $N_c-1$. This shows that all operators at the level 
of $V_+\phi_1^k(\tphi^i\phi_i)^n$ cancel in the index for $n<m=N_c-k$. 

For $n=m$, the first special case arises. This happens because in the 
tensor product of flavour representations of $\phi_i^m$ from 
$(\tphi^i\phi_i)^{N_c-k}$, and of $\phi_i^k$ from the $(\lambda_1^i\phi_i)^k$ 
in the last term in the sum over $l$, the totally antisymmetric representation vanishes. 
This leads to a non-canceling contribution from the penultimate term in 
the series, namely the totally antisymmetric representation of 
$\phi_1 \phi_i^{N_c-1}$. All other representations cancel as for $n<m$. 
Thus, the lightest surviving operator in this case has 
the $SU(N_f)_l \times SU(N_f)_r$ flavour representation
\be
\left( \nyantisymtab{N_c} \quad , \quad  \nyantisymtab{N_c-k} \right),
\ee
since the ${\tilde \phi}_i$ must also be multiplied anti-symmetrically.

For completeness we now show that the operators at the level of 
$\phi_1^{k-n}\bar{\tilde{\psi}}_1^n$ with $n>0$, which could be {\it lighter} 
than the operators considered above, actually vanish.
The index contribution at this level, for a fixed value of $n$, is given schematically by 
the sum 
\be
\sum_{l=0}^n \bar{\tpsi}_1^{n-l}(\lambda_1^i\bar{\tpsi}_i)^l 
  \left( \sum_{p=0}^{k-n} (-1)^{n-l+p} \phi_1^{k-n-p}(\lambda_1^i\phi_i)^p \right).
\ee
Notice that the summation over $p$, for a fixed value of $l$ in the outer 
summation, is exactly the Index contribution of operators at the level of 
operators $\phi_1^{k'}$ in a $U(N_c')_{k'}$ theory with 
$N_c'=N_c-l, k'=k-n$. Since $N_c'-k'=(N_c-k)+(n-l)>0$, the contribution 
of this series vanishes for each allowed value of $n\geq l\geq 0$, and hence 
the whole series vanishes including the sum over $l$.

\subsubsection{Case 2: $k\geq N_c$ and $m\equiv N_c-N_f>0$}\label{sec:ncc3}

In this case we show that the operators at the leading order occur at the level 
of  $V_+\bar{\tpsi}_1^n\phi_1^{k-n}$ for $n=m$, while for $n<m$ they all cancel. The Index 
contribution at this level for fixed 
$n>0$ is given by the following 
series
\be\label{srs3b}
\sum_{l=\max(0,n-N_f)}^{n} \bar{\tpsi}_1^{n-l}(\lambda_1^i \bar{\tpsi}_i)^l 
  \left( \sum_{p=0}^{\min(k-n,N_c-1-l,N_f)} (-1)^{n-l+p} \phi_1^{k-n-p}(\lambda_1^i\phi_i)^p \right).
\ee
Now for $n<m=N_c-N_f$ and $k\geq N_c$ we have 
\be 
\min(k-n,N_c-1-l,N_f)=N_f,
\ee
which results in the vanishing of the series in brackets for each value of $l$ 
in the outer sum, using the arguments above. 

For $n=l=m=N_c-N_f$ though, we have 
\be
\min(k-m,N_c-1-l,N_f)=N_f-1.
\ee
In this case as well the series inside the brackets in \eqref{srs3b} vanishes 
for all terms in the outer sum except the last one, $l=m$, for which 
the flavour representation is easily obtained from \eqref{ncc1rep}. Including the 
$SU(N_f)_r$ representation of the $\bar{\tpsi}$'s we get
\be 
\left( \nysymtab{k-N_c}\quad , \quad \nysymtabbar{N_c-N_f} \right).
\ee

Again for completeness we now argue that the operators at the level of 
$V_+\phi_1^k(\tphi^i\phi_i)^n$ do not contribute for all $n>0$. The Index 
contribution at this level is given by the  series:
\be
(\tphi^i\phi_i)^n \left(\sum_{l=0}^{N_f} (-1)^l \phi_1^{k-l}(\lambda_1^i\phi_i)^l \right).
\ee
Here, since $N_c>N_f$, all possible representations in the flavour tensoring of 
$\phi_i^n$ with $(\lambda_1^i\phi_i)^l$ are present, and hence the sum cancels 
identically due to the arguments above for the vanishing of 
the series inside the brackets.

\subsection{$U(N_c)_k(N_f,0)$ theories}\label{sec:ncaa}

In these ``{\it chiral}" theories, since we only have chiral multiplets in the 
fundamental representation of the gauge group, we only have as supersymmetric letters
positively charged $\phi$'s 
and negatively charged $\bpsi$'s under the Cartan of the gauge group. Thus, 
in contrast to $U(N_c)_k(N_f,N_f)$ theories, the positively charged bare 
monopole operators have to be dressed with $\phi$'s, while 
negatively charged bare monopole operators have to be dressed 
with $\bpsi$'s and are thus very different.

Another important feature of these theories is that the duality 
for these theories works differently depending on the sign of $(k-\half N_f)$:
\be\begin{split}\label{csdtab2}
& k-\half N_f \geq 0 : \quad U(N_c)_k (N_f,0) \leftrightarrows 
                          U(|k|+\half N_f-N_c)_{-k}(0,N_f), \\
& k-\half N_f \leq 0 : \quad U(N_c)_k (N_f,0) \leftrightarrows 
                          U(N_f-N_c)_{-k}(0,N_f). \\
\end{split}\ee

In the following subsections we will analytically determine the leading 
monopole operators (${\cal M}_\pm$) for a subset of the possible cases,
including all cases where the leading operator does not involve
derivatives or $(\phi_i \bar\psi^i)$ factors.

For convenience we tabulate the relevant charges of the basic supersymmetric 
letters and the bare monopole operators in these theories in Table $5$.
\begin{table}
\centering
\begin{tabular}{|c|c|c|c|}
\hline
Operator & $U(1)$ & $\epsilon +j_3$ & $SU(N_f)$ \\
\hline
$V_+$ & $-k-\half N_f$ & $\half(1-r)N_f-N_c+1$ & ${\cal I}$ \\ 
\hline
$V_-$ & $k-\half N_f$ & $\half(1-r)N_f-N_c+1$  & ${\cal I}$ \\ 
\hline
$\phi_1,~\lambda_1^i\phi_i$ & $+1$ & $1+r$ & $N_f$ \\
\hline
$\bpsi^1,~\lambda_i^1\bpsi^i$ & $-1$ & $3-r$  & $\bar{N}_f$ \\
\hline
$\phi_i \bpsi^i$ & 0 & 2 & $N_f\times \bar{N}_f$ \\
\hline
\end{tabular}
\caption{Some basic $U(N_c-1)$-invariant supersymmetric letters and their gauge and global charges in 
$U(N_c)_k(N_f,0)$ theories.}
\label{cblc}
\end{table}

\subsubsection{${\cal M}_-$ for $k>\half N_f$}

In this subsection we analyze the operators of the schematic form 
${\cal M}_-=V_-\bpsi_1^{k-\half N_f}$, possibly with derivatives sprinkled 
over the $\bpsi$'s. This case can be divided into subcases, depending 
on the comparison between $k-\half N_f$ and $N_c$.

\paragraph{$\bf{N_c-1\geq k-\half N_f>0}$ :} The Index contribution at the level of 
the lowest possible ${\cal M}_-$ is given by the following series
\be
\sum_{l=\max(0,k-\thalf N_f)}^{k-\half N_f} (-1)^l (\bpsi_1)^{k-\half N_f-l} (\lambda_1^i\bpsi_i)^l.
\ee 
Again, as in previous subsections, since the ``$x$ charge" of the operator in each term is 
the same, we can just work with the fermion number and $SU(N_f)$ representation of 
the operators. This is given by
\be\begin{split}\label{cc1rep}
 & \sum_{l=\max(0,k-\thalf N_f)}^{k-\half N_f} (-1)^l \left( \quad \nyantisymtabbar{k-\half N_f-l} 
          \otimes \nysymtabbar{l} \quad \right) \\
= & \sum_{l=\max(0,k-\thalf N_f)}^{k-\half N_f} (-1)^l\left( \nytabbar{l}{k-\half N_f-l+1} 
          \oplus \nytabbar{l+1}{k-\half N_f-l} \right)
=   0. \\
\end{split}\ee 

Since the level considered above vanishes, we have go to higher levels by sprinkling 
derivatives over the $\psi$'s, and/or attaching gauge-invariants ($\phi_i\bar\psi^i$) on 
top of the above operators. We will not perform this analysis, but in section \ref{chiral} we give a conjecture 
for the leading operators in this case (Case 1a) based on Mathematica, and show its 
consistency with the dualities discussed in 
\cite{Benini:2011mf}.

\paragraph{$\bf{ N_f+N_c-1\geq k-\half N_f\geq N_c}$ :} 

In this case (Case 1b) one of the operators at the level discussed 
in the previous case survives cancellation. The Index contribution is given by 
\be\begin{split}\label{cc2rep}
& \sum_{l=\max(0,k-\thalf N_f)}^{N_c-1} (-1)^l (\bpsi_1)^{k-\half N_f-l} (\lambda_1^i\bpsi_i)^l.  \\
= & \sum_{l=\max(0,k-\thalf N_f)}^{N_c-1} (-1)^l \left( \quad \nyantisymtabbar{k-\half N_f-l} 
          \otimes \nysymtabbar{l} \quad \right) \\
= & \sum_{l=\max(0,k-\thalf N_f)}^{N_c-1} (-1)^l\left( \nytabbar{l}{k-\half N_f-l+1} \oplus 
    \nytabbar{l+1}{k-\half N_f-l} \right) \\
= & (-1)^{N_c-1} \nytabbar{N_c}{k-\half N_f-N_c+1}.  \\
\end{split}\ee

\paragraph{$\bf{k-\half N_f=(nN_f+m)+(N_c-1) \  with~ n\geq 1 ~and~ N_f>m\geq 0 }$ :} 

In this case (Case 1c) there is a unique operator present at the minimal level,
\be 
{\cal M}_{-}= V_{-}(\lambda_1^i\bpsi_i)^{N_c-1} (\bpsi_1)^{N_f} 
      (\del\bpsi_1)^{N_f} \ldots (\del^{n-1}\bpsi_1)^{N_f} (\del^n\bpsi_1)^m .
\ee
Notice that in this operator none of the $\bpsi_1$'s can be replaced with 
$\lambda_1^i\bpsi_i$, as the resulting operator would vanish due to antisymmetry of 
more than $(N_c-1)$ $\lambda_1^i$'s. Furthermore, none of the $\del$'s can be replaced 
with $\phi^i\bpsi_i$, as the resulting operators would vanish due to antisymmetry of 
more then $N_f$ $\bpsi_1$'s. This proves that this is the unique leading operator in this case.

To determine the flavour representation of this operator note that each of the 
$(\del^l\bpsi_1)^{N_f}$ factors forms a flavour singlet, while the remaining factors 
give the representation
\be 
 \left( \nysymtabbar{N_c-1}\otimes \nyantisymtabbar{m} \right).
\ee

\subsubsection{${\cal M}_-$ for $k<\half N_f$}

Since in this case $k-\half N_f<0$, we need to dress $V_{-}$ by 
$\phi_1$'s (as opposed to $\bpsi_1$'s in the previous cases) to make it gauge-invariant. 
Schematically we have 
\be  \label{monopoleone}
{\cal M}_-= V_- (\phi_1)^{\half N_f-k}.
\ee
A straightforward application of the arguments presented in section \ref{sec:ncc1} gives
us the following results for the operator of this form contributing to the index in this case:
\be\begin{split}\label{oncmm}
{\cal M}_- &= 0 \quad \text{ for } \quad \half N_f-k<N_c, \\
{\cal M}_- &=  \nytab{\half N_f-k-N_c+1}{N_c} \quad \text{ for } 
                       \quad \half N_f-k \geq N_c. \\
\end{split}\ee
The second case is Case 2a, for which we have found the leading monopole operator.
For the first case (Case 2b) we need to add derivatives and/or gauge-invariants 
on top of the operator \eqref{monopoleone}. We will not do this here, but we give a general conjecture based on 
results obtained using Mathematica for low values of $(k,N_c,N_f)$ in Table \ref{Chiralminus}.

\subsubsection{${\cal M}_+$ for $k>\half N_f$}
From Table $5$ we see that the naive lowest ${\cal M}_+$ in this case 
is of the schematic form $V_+ (\phi_1)^{k+\half N_f}$. A straightforward 
application of the arguments presented above gives the contribution 
at this level
\be\label{ccphi}
{\cal M}_{+} =  \nytab{k+\half N_f-N_c+1}{N_c} 
                      \quad \text{for} ~~N_c\leq  N_f, ~k+\half N_f-N_c\geq 0.
\ee
Notice that the second condition in \eqref{ccphi} above is the same as the condition for 
the existence of a supersymmetric vacuum in these theories. Thus, within the set of theories 
possessing a supersymmetric vacuum, this level survives in the Index for 
$N_f\geq N_c$ (Case 1a, as we saw in section \ref{analyticone}). 

For $N_f<N_c$ (Cases 1b and 1c), this level vanishes and we need to 
consider operators with derivatives and/or gauge invariants ($\phi^i\bpsi_i$). 
The analytic analysis for this gets complicated and we will not pursue it here. 
Instead we present a conjecture for these cases in Table \ref{Chiralplus},
based on Mathematica evaluations at low values of $k,N_c$ and $N_f$.

\subsubsection{${\cal M}_+$  for $k<\half N_f$}

For $k<\half N_f$, the condition for the existence of a supersymmetric vacuum 
is $N_f\geq N_c$. Thus the condition for the $V_+ (\phi_1)^{k+\half N_f}$ level 
to survive is $k+\half N_f-N_c \geq 0$ in \eqref{ccphi} (this is Case 2b that we
analyzed already in section \ref{analyticone}). For $k+\half N_f-N_c<0$ (Case 2a), this level 
vanishes in the Index and we need to consider operators with derivatives and/or gauge-invariants ($\phi^i\bpsi_i$). Again, we will not pursue this exercise here, but present a 
conjecture in Table \ref{Chiralplus}, based on Mathematica 
evaluations for low values of the parameters.

\section{Dimensions of the lowest monopole operators in a Chern-Simons-fermion theory}
\label{fermion_counting}

In this section we discuss the lowest monopole operators in (non-supersymmetric) theories of fermions in the fundamental representation
coupled to a $U(N_c)_k$ Chern-Simons theory. For simplicity we focus on the case of a single flavour, for which $k$ must be half-integer.
As mentioned in the introduction, in a theory with only fundamental fermions, we expect the lowest monopole operator to arise from a product of a bare $(1,\vec{0})$ monopole operator with $|k|-\half$ fermions.  In this appendix we will compute the naive scaling dimension of such an operator. We are
mostly interested in how this dimension scales for large $|k|$.

The main point is that in the case of fermions, because of Fermi statistics, one cannot just add $\psi^{k-\half}$ 
to a bare monopole operator. One necessarily has to include fermions dressed with derivatives to construct a product with more than two fermions.
If we needed to construct an operator of the form $\psi^{k-\half}$ without the monopole background (ignoring the fact that this would not be gauge-invariant), we would use the fact that  the fermion operators with $n$ derivatives $D_{\alpha_1 \beta_1}\cdots D_{\alpha_n \beta_n} \psi_\alpha$ form a spin $(n+{1 \over 2})$ representation\footnote{Antisymmetrizations between derivative and fermion indices are removed by the equation of motion which fixes $\epsilon^{\beta \gamma}D_{\alpha  \beta } \psi_\gamma$. Spin singlet derivatives $D^2 \psi_\alpha$ vanish for the same reason. Since the derivatives commute, the only remaining representation is the symmetrized product of fermions and derivatives that has spin $n + {1 \over 2}$.}. Hence their number is given by $2 (n+{1 \over 2})+1 = 2n +2$. Thus the schematic operator is
\begin{equation} \label{fermmon}
(\psi)^2 \cdots (D^n \psi)^{2n +2}\cdots. 
\end{equation}
We see that in an operator with order $k$ fermions, we must have factors $D^{n_{max}} \psi$ with $n_{max}$ at least of order $\sqrt{k}$.
The total number of derivatives acting on all $k$ fermions in such an operator is then at least of order $n_{max}^3 \sim k^{3 \over 2}$. Each operator $D^n \psi$ has classical dimension $(n+1)$. Hence, the naive scaling dimension of such an operator is $O(k^{3 \over 2})$.   

The only modification in the monopole background is that now the spectrum of fermions is shifted down by a half, namely they have spins $n=0,1,2,\cdots$,
with the energy on $S^2$ equal to the spin \cite{Wu:1976ge}. First, this means that there is now a fermionic zero mode, so there are two
bare monopole operators with the lowest dimension and with charges $k \pm \half$. Second, this means that the product in \eqref{fermmon}
involves operators with multiplicity $2n+1$ for $n=1,2,\cdots$. However, this does not modify the analysis above for large $k$, so we still
find that the dimension of the lowest monopole operator is naively of order $k^{3\over 2}$.

As discussed in section \ref{nspc}, these naive dimensions may have large corrections that we do not know how to control.


\section{Additional charges of monopoles and their matching}\label{apdx:DCM}
In this appendix we give the global charges of the chiral monopole operators presented in sections \ref{nonchiral} and \ref{chiral}, beyond the $SU(N_f)$ charges discussed there.  We also show that the global charges of the dual operators match across the GK duality. We have already presented one combination of the global charges $2 j_3 + R$, appearing in the index, in sections \ref{nonchiral} and \ref{chiral}. 

Let us start with the nonchiral case. The global charges are $(j_3,R ,A)$ as mentioned in section \ref{background}. For the monopole operators listed in Table \ref{NonChiral}, we give these global charges in Table \ref{NonChiralGlobal}. Here $N_c^d = k + N_f - N_c$ is the rank of the dual group. 
Note that we cannot read off the $j_3$ and $R$-charges just from the index. For Cases 1, 2 and 4, the value of $j_3$ is computed from the form of the
leading monopole operator that we found analytically. One can verify that the charges of Cases 2 and 4 map correctly under the duality.   For
Case 3 we use the operator that we conjectured in Table \ref{NonChiral}, and one can check that this is consistent with the duality to  Case 1. For completeness, we mention that for $k=0$, the naive chiral operator $V_+$ survives, whose charges are  $(j_3,R,A) =(0,N_f(1-r) - N_c +1,-N_f)$.  We also note the charges of the superconformal primary from which the corresponding monopole operator descends in Table \ref{NonChiralGlobalSupConf}, using the rules mentioned in \cite{Minwalla:2011ma}.

\begin{table}
\centering
\begin{tabular}{|c|c|c|}
\hline
  & Region & ($j_3,R,A$)  \\
\hline
1 &  $N_f > N_c , k \ge N_c $  &  $\left( { k-N_c+1 \over 2}, \ N_f+r(k-N_f), \ k - N_f \right) $  \\
\hline
2 &  $k \ge N_c \ge N_f$  &  $\left({ k-N_f+1\over 2 }, \ N_c(1-r)+N_c^d r , \ N_c^d - N_c\right)$ \\
\hline
3 &    $N_c \ge N_f , N_c >  k$  &  $ \left({ N_c-N_f+1 \over 2 }, \ k+r(N_f-k),\  N_f -k \right) $ \\
\hline
4 & $ N_f \ge N_c \ge k $  &  $\left(0, \ N_c^d(1-r)+rN_c+1 ,\  N_c - N_c^d \right)$  \\
\hline
\end{tabular}
\caption{Global charges of the leading ${\cal M}_+$ monopole operators in $U(N_c)_k(N_f,N_f)$ theories in various parameter ranges.}
\label{NonChiralGlobal}
\end{table}

\begin{table}
\centering
\begin{tabular}{|c|c|c|}
\hline
  & Region & ${\cal N} = 2$ primary charges:  $\epsilon$, ($j_3,R,A$)  \\
\hline
1 &  $N_f > N_c , k \ge N_c $  &  $\epsilon = j_3+R+1$\ ,\ $\left( { k-N_c \over 2}, \ N_f-1+r(k-N_f), \ k - N_f \right) $  \\
\hline
2 &  $k \ge N_c \ge N_f$  & $\epsilon = j_3+R+1$\ ,\  $\left({ k-N_f\over 2 }, \ N_c(1-r)+N_c^d r-1 , \ N_c^d - N_c\right)$ \\
\hline
3 &    $N_c \ge N_f , N_c >  k$  & $\epsilon = j_3+R+1$\ ,\    $ \left(  { N_c-N_f \over 2 }, \ k+r(N_f-k)-1,\  N_f -k \right) $ \\
\hline
4 & $ N_f \ge N_c \ge k $  &$\epsilon = R $\ ,\    $\left( 0, \ N_c^d(1-r)+rN_c+1 ,\  N_c - N_c^d \right)$  \\
\hline
\end{tabular}
\caption{Global charges of the ${\cal N} =2$ superconformal primary corresponding to the ${\cal M}_+$ monopole operator in $U(N_c)_k(N_f,N_f)$ theories in various parameter ranges.}
\label{NonChiralGlobalSupConf}
\end{table}


Let us now consider the chiral case. Some of the charges of the two monopole operators $({\cal M}_+, { \cal M}_-)$ which survive in this case were given in Table \ref{Chiralplus} and Table \ref{Chiralminus}. With the same conventions for regions, we give the global charges $(j_3,R)$ in Table \ref{ChiralGlobal}.
Again, in the cases in which we computed the monopole operator explicitly, the charge we give is based on this operator. In the other
cases the charge we give is based on our conjectures in Tables \ref{Chiralplus} and \ref{Chiralminus}, and one can verify that in
all cases it is consistent with the duality. We also note the charges of the superconformal primary from which the corresponding monopole operator descends in Table \ref{ChiralGlobalSupConf}, using the rules mentioned in \cite{Minwalla:2011ma}.

\begin{table}
\centering
\begin{tabular}{|c|c|c|}
\hline
 & ${ \cal M}_+$ charges $(j_3,R)$  &  ${\cal M}_-$ charges $(j_3,R)$ \\
\hline
1a &  $\left( { \tilde  N_c +1 \over 2 } , {N_f \over 2} + k r \right) $ & 
$\left( { N_c+1  \over 2} , k (1-r) \right) $  \\
\hline
1b &  $\left( { k+N_c-\thalf N_f+1 \over 2}, \half N_f+kr \right)$  & 
$\left( k-\half N_f-\half N_c+\half, k(1-r) \right)$ \\
\hline
1c &  $\left(\frac{N_c^d-1}{2}+\frac{n(n+1)}{2}N_f+m(n+1),\half N_f+kr \right)$ & 
$\left(\frac{N_c-1}{2}+\frac{n(n+1)}{2}N_f+m(n+1),k(1-r) \right)$ \\
\hline
2a & $\left( { N_c-k-\half N_f+1  \over 2 }, \half N_f+kr \right)$  & 
$\left({ \half N_f-k-N_c+1 \over 2} , \half N_f-kr \right)$  \\
\hline
2b  & $ \left( { k+\half N_f-N_c+1 \over 2} , \half N_f+kr \right) $  &  
$\left( { k+N_c-\half N_f+1 \over 2 }, \half N_f-kr \right)$  \\
\hline
\end{tabular}
\caption{Global charges of the leading ${\cal M}_+, {\cal M}_-$ monopole operators in $U(N_c)_k(N_f,0)$ theories.}
\label{ChiralGlobal}
\end{table}

\begin{table}
\centering
\begin{tabular}{|c|c|c|}
\hline
 & ${ \cal N}=2$ primary $(j_3,R)$; $\epsilon = j_3 + R + 1$  &  ${ \cal N}=2$ primary charges $(j_3,R)$; $\epsilon = j_3 + R + 1$\\
\hline
1a &  $\left( { \tilde  N_c   \over 2 } , {N_f \over 2} + k r -1\right) $ & 
$\left( { N_c   \over 2} , k (1-r)-1 \right) $  \\
\hline
1b &  $\left( { k+N_c-\thalf N_f  \over 2},{N_f \over 2}+kr-1 \right)$  & 
$\left( k-\half N_f-\half N_c , k(1-r)-1 \right)$ \\
\hline
1c &  $\left(\frac{N_c^d-2}{2}+\frac{n(n+1)}{2}N_f+m(n+1) ,{N_f-2 \over 2}+kr \right)$ & 
$\left(\frac{N_c-2}{2}+\frac{n(n+1)}{2}N_f+m(n+1) ,k(1-r) -1\right)$ \\
\hline
2a & $\left( { N_c-k-\half N_f   \over 2 }, {N_f \over 2}+kr -1\right)$  & 
$\left({ \half N_f-k-N_c \over 2} , {N_f \over 2}-kr -1\right)$  \\
\hline
2b  & $ \left( { k+\half N_f-N_c  \over 2} , {N_f \over 2}+kr-1 \right) $  &  
$\left( { k+N_c-\half N_f  \over 2 },{N_f \over 2}-kr -1\right)$  \\
\hline
\end{tabular}
\caption{Global charges of the ${\cal N} =2$ superconformal primary corresponding to the ${\cal M}_+, {\cal M}_-$ monopole operators in $U(N_c)_k(N_f,0)$ theories in various parameter ranges.}
\label{ChiralGlobalSupConf}
\end{table}

Note that $j_3$ matches straightforwardly under the duality, while for the $R$ charge one needs 
an extra shift by  $-{N_f \over 2}$ for Cases 1a,1b,1c, and by $-k$ in Cases 2a,2b, as explicitly 
given in (\ref{ChiralMatch})\footnote{These shifts in R-charge, required for duality matching, 
are from the side of the original theory. The corresponding shifts from the dual side are just the negatives of 
these.}.

As discussed in the main text, $k=\half N_f$ is a special case and to find the 
lightest ${\cal M}_+$  monopole operator on the dual side one needs to take into account the 
contribution of the gauge singlet chiral multiplet  $V_+$. Here we present the 
charges of the operator in \eqref{cvpd} of the dual $U(N_c^d)_{-\half N_f}(0,N_f)$ 
theory, which are
\be\label{cscc}
(\half(N_c^d+1),N_f(1-r)).
\ee
With a shift of $\half N_f$ in the R-charge these match precisely with those of ${\cal M}_+$ 
in Case 1a (or 2b) for $k=\half N_f$. Further it is easily verified that our proposed dual 
operator \eqref{cvpd} contains, in its flavour decomposition, the flavour representation 
of the corresponding operator in the original theory.

\bibliographystyle{JHEP}
\bibliography{monopole}

\end{document}